\documentclass[useAMS,usenatbib]{mn2e}

\usepackage{amssymb}
\usepackage{amsmath}
\usepackage{graphicx}
\usepackage{BibDef}
\usepackage{hyperref}
\usepackage{subfigure}

\usepackage{breakurl}
\def\lsim{\mathrel{\rlap{\lower 3pt \hbox{$\sim$}} \raise 2.0pt \hbox{$<$}}}
\def\gsim{\mathrel{\rlap{\lower 3pt \hbox{$\sim$}} \raise 2.0pt \hbox{$>$}}}
\def\msun{\rm {M_{\large \odot}}}

\title[MBH dynamics in AMR simulations] {Massive black hole and gas dynamics in galaxy nuclei mergers. I. Numerical implementation}
\author[Lupi et al.]{Alessandro Lupi $^{1}$, Francesco Haardt$^{1,3}$ \& Massimo Dotti$^{2,3}$.\\
$^1$DiSAT, Universit\`a degli Studi dell'Insubria, Via Valleggio 11, 22100 Como, Italy\\
$^2$Dipartimento di Fisica, Universit\`a degli Studi di Milano-Bicocca, Piazza della Scienza 3, Milano 20126, Italy\\
$^3$INFN, Sezione di Milano-Bicocca, Piazza della Scienza 3, 20126 Milano, Italy}
\begin{document}

\date{Draft July 2014}

\pagerange{\pageref{firstpage}--\pageref{lastpage}} \pubyear{2013}

\maketitle

\label{firstpage}

\begin{abstract} 
Numerical effects are known to plague adaptive mesh refinement (AMR) codes
when treating massive particles, e.g. representing massive black holes (MBHs). 
In an evolving background, they can experience strong, spurious perturbations and then follow unphysical orbits. 
We study by means of numerical simulations the dynamical evolution of a pair MBHs in the rapidly and violently evolving gaseous and stellar background that follows a galaxy major merger. 
We confirm that spurious numerical effects alter the MBH orbits in AMR simulations, 
and show that numerical issues are ultimately due to a drop in the spatial resolution during the simulation,  
drastically reducing the accuracy in the gravitational force computation.  
We therefore propose a new refinement criterion suited for massive particles, 
able to solve in a fast and precise way for their orbits in highly dynamical backgrounds. 
The new refinement criterion we designed enforces the
region around each massive particle to remain at the maximum
resolution allowed, independently upon the local gas density. Such maximally-resolved regions then follow the MBHs along their orbits, and effectively avoids all spurious effects caused by resolution changes. 
Our suite of high resolution, adaptive mesh-refinement hydrodynamic simulations, including different 
prescriptions for the sub-grid gas physics, shows that the new refinement implementation has the advantage of not altering the
physical evolution of the MBHs, accounting for all the non trivial physical processes taking place in
violent dynamical scenarios, such as the final stages of a galaxy major merger.

\end{abstract}

\section{Introduction}
\label{sec:Intro}
Mergers of gas--rich galaxies are key events in hierarchical clustering models of cosmic structure formation.     
Galaxy mergers are thought to $(i)$ be responsible of the formation of classical bulges
\citep[e.g.][and references therein]{kormendy13}, $(ii)$ trigger
intense bursts of nuclear star formation and AGN emission
\citep[e.g.][]{dimatteo05,daddi10, genzel10} with possible dual AGN
activity \citep[e.g.][]{callegari09, callegari11, vanwassenhove12,
  blecha13}, $(iii)$
trigger the formation of massive black holes (MBH), possibly via direct collapse of gas
\citep[][]{mayer10, bonoli14}, or inducing the collapse of
pre-existing nuclear stellar clusters \citep{davies11,lupi14}, and 
$(iv)$ build up a population of MBH binaries (MBHBs)
\citep[e.g.][and references therein for a detailed
  discussion]{colpi11}, the main targets for future campaigns aimed at the detection of low--frequency gravitational waves \citep{lisa, PTA}.  

When simulating MBHs in galaxy mergers, 
a detailed treatment of the dynamics is of foremost importance, since it influences the MBH
ability to accrete gas (hence, its mass evolution and possible onset of AGN activity), the MBH spin evolution \citep[e.g.][]{dotti10},
and the formation and fate of MBHBs. The many physical processes involved require a detailed modelling of the dynamics of gas, stars, dark
matter and MBHs though, hence calling for high--resolution hydrodynamical simulations. 

Hydro simulations can be divided into two classes:
smoothed particle hydrodynamics (SPH), in which the distribution of gas is
sampled by a number of particles, and grid codes, in which the gas 
properties are mapped on a geometrical grid. SPH codes do not assume any a
priori geometry, and automatically allow larger resolution
in denser regions. Therefore, SPH codes are extensively
used in galaxy merger simulations \citep[see][and references 
therein]{dotti12}. 

Grid codes, in order to achieve high resolution only in a limited volume of the
simulation, can refine the mesh when and where
required by specified properties of the fluid \citep[adaptive mesh refinement, AMR][]{berger84,
berger89}. As an example, AMR allows to
evolve the dynamics of gas close to MBHs, and then, e.g., determine whether AGN
activity is to be expected or not, without increasing dramatically the
total numbers of grids and the computational cost of the run. In
principle, one can tailor the refinement criteria to obtain a
super--Lagrangian increase on the resolution, allowing, e.g., for a
better description of MBH sinking in late stages of galaxy mergers
\citep[see][]{chapon13}. Moreover, the resolution of the gravitational
interaction depends on the degree of refinement, and can change and
increase as the simulation proceeds, a feature not generally shared with SPH codes.

Thanks to AMR, a number of grid--based hydro simulations of the last
stages of galaxy mergers have been performed to date
\citep[][]{chapon13, dubois14a}. The two papers assume quite
  different prescriptions. \cite{chapon13} assume a smoother IGM, not
  affected by cooling, star formation (SF) and supernova (SNa)
  feedback, while these effects are considered in \cite{dubois14a}. In
  \cite{chapon13} the MBH dynamics depends strongly on the maximal
  resolution of the simulation. In lower resolution runs ($\Delta
  x_{\rm min}=3$ pc) the MBH evolution is significantly slower
  (because of the underestimated effect of the resolution dependent
  dynamical friction), and considerably more noisy (well above the
  resolution level) than in the higher resolution cases ($\Delta
  x_{\rm min}=0.1$ pc). A similar noisy evolution of the MBH orbits
has been observed by \cite{dubois14a}. In order to prevent spurious
oscillations of the MBH due to finite resolution effects, the authors
introduced an additional drag force onto the MBHs.

Interestingly, a noisy evolution of collisionless particle dynamics
(and, in particular, of MBHs) has been observed in high--resolution
AMR simulations of single isolated galaxies, in which the gas is only
subject to internal processes such as star formation, SNa feedback,
etc. \citep[e.g.][]{gabor13}. In their work, in order to limit
numerical MBH wandering, the authors propose two different
approaches. The first one consists in modelling the MBH as an extend
spherical structure, using few thousands evolving particles. Such
BH--forming particles are regenerated over a secondary, coarse time
grid.  In this case the MBH moves out of the geometrical centre of the
galaxy by hundreds to thousands of pc depending on the amount of gas
simulated and the noisy effect is only reduced.  The second one,
instead, consists in adding an artificial velocity component directed
toward the stellar centre of mass, which forces the MBH to orbit close
to the galaxy centre.

As noted by \cite{gabor13}, the noisy evolution of the MBH can be
either numerical (due to the limited and time varying spatial
resolution), or physical, if caused by interactions with massive and
dense gas clouds. This last possibility is particularly interesting
when the gas is allowed to cool and actually form significant compact
overdensities, as in the simulations discussed in \cite{gabor13} and
\cite{dubois14a}. Indeed, a physically motivated noisy orbital
evolution of MBHs is observed in SPH simulations \citep[see e.g.][for
  a detailed and extensive discussion]{fiacconi13}. It is important to
notice that the effects of massive gas clumps on the MBH dynamics is
severely altered by the corrections proposed in works discussed above.

This is the first of a series of papers devoted at the study of MBHB
dynamics in the final stages of galaxy mergers.  Here, we focus on the
technical aspects of numerical simulations of the physical system
under scrutiny, and propose a new adaptive refinement criterion for
AMR codes, suited to properly treat the physical interactions between
the MBHs and the gas clouds forming in the environment.  In
forthcoming papers we will exploit our newly designed simulations,
exploring the parameters space, and addressing the many diverse
astrophysical and cosmological consequences.

The paper is organised as follows. In Section~2 we introduce the code
we developed to build our initial conditions, the simulations we have
ran, and our new implementation of an evolving geometrical refinement
that we implemented on the AMR code RAMSES \citep{teyssier02}, in
Section~3 we present the results of our test runs and compare them to
results obtained without implementing our new refinement
criterion. The conclusions are drawn in Section~4.

\section{Numerical methods}
\label{sec:ic}
\subsection{Initial conditions}

We simulated the merger of two circum--nuclear discs (CNDs) embedded in stellar nuclei, each hosting a MBH.  Such system should be considered as an idealised model of the latest stages of a galaxy merger. We initially set  each of the two merging nuclei in dynamical equilibrium, and assumed they are constituted by three different components:
\begin{itemize}
\item a stellar spherical structure (termed ``nucleus'' hereinafter) described by an Hernquist profile \citep{hernquist90}, defined in spherical coordinates as
\begin{equation}
\rho_b(r)=\frac{M_b}{2\pi}\frac{a}{r\left(r+a\right)^3},
\end{equation}
where $\rho_b(r)$ is the density as a function of radius  $r$, $M_b=2\times 10^8 \msun$ the total nucleus mass, and $a=100$ pc the nucleus scale radius.
\item an exponential gaseous disc with surface density profile defined in cylindrical coordinates as 
\begin{equation}
\Sigma_d(R)=\frac{M_d}{2\pi R_d^2} \exp(-R/R_d),
\end{equation}
where $R$ is the disc radius, $R_d= 50$ pc the disc scale radius, and $M_d=10^8 \msun$ the total disc mass.
\item a MBH with mass $M_{\rm BH}=10^7 \msun$, at rest in the centre of the disc.
\end{itemize}

In order to ensure equilibrium, we iteratively calculated the vertical
density profile of the disc and, consequently, the velocity fields of
gas and stars by means of the dedicated code GD\_BASIC\footnote{The
  code is publicly available at
  \url{http://www.dfm.uninsubria.it/alupi/software.html}}.  The code
samples gaseous and stellar particles, eventually used as the initial
condition for the SPH run (see next section~\ref{sec:suite}).  GD\_BASIC solves the
disc hydrostatic equilibrium equations for a user-defined surface
density profile subject to the additional potentials of the Hernquist
nucleus and of the MBH. Defining the gas pressure as
\begin{equation}
P_d=(\gamma-1)\rho_d u,
\label{eq:poly} 
\end{equation}
where $\gamma$ and $u$ are the gas polytropic index and internal energy respectively, 
and assuming a single temperature disc (in our case, we adopted an initial fiducial value of $T=2\times 10^4$ K), the vertical equilibrium equation can then be written as
\begin{equation}
\frac{1}{\rho_d(R,z)}\frac{\partial P_d(R,z)}{\partial z}=-\frac{\partial \phi(R,z)}{\partial z},
\label{eq:vs}
\end{equation}
where $\phi(R,z)$ is the global gravitational potential of the system.
Using eq.~\ref{eq:poly}, eq.~\ref{eq:vs} can be solved for the disc surface density: 
\begin{equation}
\Sigma_d(R)=\rho_d(R,0)\int_{-\infty}^{+\infty}\exp\left[-\frac{\phi_z(R,z)}{(\gamma-1)u}\right]dz,
\label{eq:rhoz}
\end{equation}
where $\phi_z(R,z) \equiv \phi(R,z)-\phi(R,0)$ is the vertical component of the global potential.

We further assumed the thin disc approximation, so that the Poisson equation can be simplified to 
\begin{equation}
\frac{\partial^2\phi_d}{\partial z^2} +\nabla^2 \phi_b =4\pi G[\rho_d(R,z)+\rho_b(r)],
\end{equation}
where $\phi_d$ and $\phi_b$ are the disc and the nucleus potentials, respectively. 
Since $\nabla^2 \phi_b=4\pi G\rho_b$, we can write 
\begin{equation}
\frac{\partial^2\phi_{d,z}}{\partial z^2} =4\pi G\rho_d(R,z)=4\pi G \rho_d(R,0)\exp\left[-\frac{\phi_z(R,z)}{(\gamma-1)u}\right],
\label{eq:poisson}
\end{equation}
where we defined $\phi_{d,z}\equiv \phi_d(R,z)-\phi_d(R,0)$ as the vertical component of the disc potential. 

We solved the above equations forcing the vertical hydrostatic equilibrium of the disc, assuming an exponential surface density profile as boundary condition. 
We started solving eq.~\ref{eq:poisson} for $\phi_{d,z}$ by guessing an initial equatorial profile $\rho_d(R,0)$. 
Then, we computed the total vertical potential $\phi_z$, and by means of eq.~\ref{eq:rhoz}, a new value for $\rho_d(R,0)$ 
that satisfies the boundary condition $\Sigma_d$ was derived. The procedure was iterated until convergence. From eq.~\ref{eq:poisson} we obtained 
$\phi_{d,z}$ and, from eq.~\ref{eq:vs}, $\rho_d(R,z)$. As in the iterative procedure $\phi_d(R,0)$ is a free parameter, we assumed 
a razor thin exponential disc (equation 2-168 in \citeauthor{binney08}). Finally, the velocity of the disc particles was evaluated by setting the radial component of the velocity equal to 0 (hydrostatic equilibrium assures that the vertical component is null as well), while the tangential velocity was obtained from the Euler equation in the case of a rotationally supported disc.

Concerning stellar-like particles, we evaluated the distribution function $f$ in the 6-dimensional phase-space. We initially considered the 
Hernquist spherical structure subject only to its own potential and to the MBH potential 
(i.e., $\phi=\phi_b+\phi_{\rm MBH}$), implying that $f$ depends only on the particle total energy in this case. 
From the Eddington's formula (equation 4-140a in \citeauthor{binney08}) we have
\begin{equation}
f(\epsilon)=\frac{1}{2^{3/2}\pi^2}\frac{d}{d\epsilon}\int_\epsilon^0 \frac{d\rho_b}{d\phi}\frac{d\phi}{(\phi-\epsilon)^{1/2}},
\end{equation}
where $\epsilon$ is the particle energy per unit of mass, and $\rho_b$ can be expressed as a function of $\phi$ (being $\phi$ a monotonic function of $r$).
The distribution function was numerically evaluated and used to sample the nucleus particle energy density.
We then derived the particle speed $v=\sqrt{2(\epsilon-\phi)}$, where $\phi$ was computed at the particle position.
In order to correct for the neglected contribution of the disc to the global potential, we added to the spherically symmetric component of the potential the approximate contribution of the disc in the form 
$\phi_d=GM_d(<r)/(3r)$, where $M_d(<r)$ is the mass of gas particles within $r$.

\begin{figure}
\includegraphics[width=0.45\textwidth]{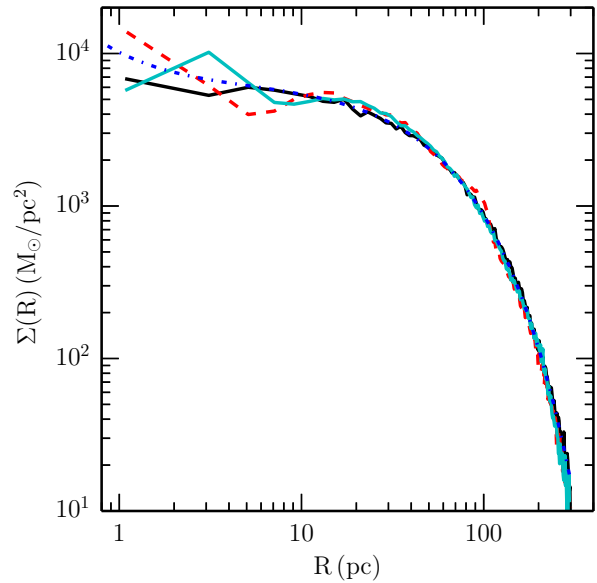}
\caption{\small Surface density profile for a disc evolved in isolation using Gadget-2. The solid black, dashed red and solid cyan curves are obtained from the gas particle distribution at $t=0, 5 $ and $10$ Myr, respectively. The dash--dotted blue curve is the profile calculated with the algorithm described in the text.}
\label{fig:surfEvol}
\end{figure}

\begin{figure}
\includegraphics[width=0.45\textwidth]{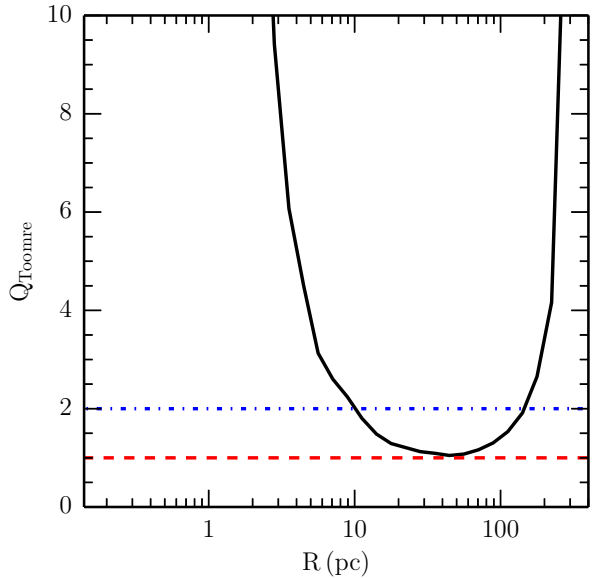}
\caption{\small Toomre parameter  at $t=0$ Myr for the isolated disc using the thin disc approximation. The dashed red line corresponds to $Q=1$, while the dash--dotted blue one to $Q=2$. The y-axis has been limited in the interval between 0 and 10 to highlight the instability interval, corresponding to $Q<2$. The region within the inner 1 pc is not visible because of the very large values assumed by Q, outside the axis limits considered.} 
\label{fig:surfToomre}
\end{figure}

In order to test the stability of our initial conditions, we evolved each disc
in isolation for 10 Myr. The evolved disc surface density is shown in
fig.~\ref{fig:surfEvol} at different times. The profile changes in the inner
$\simeq20$ pc because of a gas instability developing after $\simeq 2$ Myr
from the start. In order to assess the origin of such instability, we
numerically estimated the Toomre parameter of the disc $Q$ at initial time
$t=0$ ($Q\equiv kc_s/(\pi\Sigma)$, where $k$ is the epicyclic frequency and
$c_s$ is the gas sound speed). Note that, strictly speaking, the initialised
disc is not infinitesimally thin, so that $Q$ as defined above represents a
lower limit.  Fig.~\ref{fig:surfToomre} shows the Toomre parameter at $t=0$ as
a function of the radial distance $R$. We found $Q>2$ everywhere, with the
notable exception of the region $10\lsim R \lsim 150$ pc, where $1\lsim Q
\lsim 2$. The formation of transient spiral arms in this region, clearly seen
during the disc evolution, suggests a genuinely physical origin of the disc
instability. Such instability results in small changes in the surface density
profile in the $10\lsim R \lsim 150$ pc region. The system, now slightly out
of equilibrium, undergoes a re-adjustment of the gas distribution down to the
very central region of the disc, as observable in fig.~\ref{fig:surfEvol} down
to 5--10 pc from the MBH. Finally, no evidence of any fragmentation
instability during the overall evolution was seen, in agreement with $Q$ being
always $\gsim 1$.


\subsection{Simulation suite}
\label{sec:suite}
The procedure described above was then used to build two equal mass
co-rotating discs, each described by $10^5$ particles, at an initial
separation of $300$ pc. The initial conditions for the AMR runs have been
obtained mapping the gas particle distribution on the grid using the publicly
available code TIPGRID\footnote{The code is available at \url{http://www.astrosim.net/code/doku.php?id=home:code:analysistools:misctools}}. The discs were set on an initially
elliptical orbit with eccentricity $e=0.3$ and orbital angular momentum
antiparallel to the angular momentum of each disc. In order to test numerical
stability and assess the reliability of the system evolution, we decided to
run our simulation using two intrinsically different methodologies, using the
SPH code GADGET2 \citep{Springel05} and the AMR code RAMSES
\citep{teyssier02}, in their publicly available releases.  We note that
GADGET2 public version does not include gas cooling, supernova feedback and
star formation, while the gas obeys a polytropic equation of state. On the
contrary, the available RAMSES release includes all the mentioned physical
processes\citep{teyssier02,rasera06,teyssier13}.

As described in \S\ref{sec:Intro}, several past AMR simulations similar to what we present here are reported to show a noisy evolution of the orbits of the two 
MBHs, and different techniques have been proposed to solve the problem \citep{gabor13,dubois14a}. The reason behind numerical perturbations in the MBH motion lies most probably in the change of  the gravitational force computation accuracy during the simulation, which is related to the change in the cell size.
We therefore developed a new refinement criterion aimed at ensuring a fixed accuracy when computing the gravitational force acting on the two MBHs. We implemented a new refinement criterion based on the identity and positions of selected particles, rather than on 
the global geometry of the system. In our new implementation, refined grids
follow the positions of the two MBHs at each time-step. Surrounding cells within two specified, MBH-centered volumes are flagged for
further multi-level refinement. Up to seven concentric regions of increasing
resolution can be user-defined by setting the seven corresponding radii. For
example, in the runs discussed in this paper we enforced the maximum level of
refinement, with single cell linear sizes of 0.39 pc, within 10 pc from each
MBH. At larger distances from the MBHs the resolution degrades smoothly unless
another refinement criterion is matched. On the top of the newly implemented
criterion discussed, we use the standard Quasi-Lagrangian and Jeans criteria
already implemented in RAMSES~\citep{teyssier02}, as described below.

We performed a total of five simulations with RAMSES, in order to compare the
new refinement criterion with the standard one, and to test the reliability of
the dynamical evolution of the MBH binary under different assumptions
regarding the so-called ``sub-grid physics''.  We also carried out a single
simulation with GADGET2, assuming a polytropic equation of state with index
$\gamma=7/5$.  The spatial resolution for the SPH simulation is $0.2$ pc,
while the mass resolution is $10^3 \msun$ and $2\times 10^3 \msun$ for gas and
collisionless particles, respectively. The maximum spatial resolution (at the
highest refinement level) for all our AMR simulations is $\sim 0.39$ pc and
the mass resolution for stars is equal to that in the SPH run.  The standard
Jeans criterion enforced in all the simulations (with and without our new
refinement implementation) ensures that the Jeans length is resolved with at
least 4 cells (14 in the highest refinement level) everywhere, so to avoid the
formation of spurious clumps due to resolution limits. The Quasi-Lagrangian
criterion, on the other hand, allows us to resolve a minimum gas mass of $10^3
\msun$ everywhere, equal to the gas particle mass used in the SPH run."
\begin{table}
\centering
\begin{tabular}{l|c|c|c|c|c}
& & Star & SNa & New \\
Run & Cooling & Formation & feedback & Refinement\\
\hline
\hline
Plain & No & No & No & No\\
Plain+ & No & No & No & Yes\\
Noblast & Yes & Yes & Yes & No\\
Blast & Yes & Yes & Yes & No\\
Blast+ & Yes & Yes & Yes & Yes\\
\hline
\hline
\end{tabular}
\caption{\emph{The suite of RAMSES runs.}}
\label{tab:runs}
\end{table}
Table \ref{tab:runs} shows the suite of RAMSES simulations with the main features highlighted.  

In runs ``Plain'' and ``Plain+'' no sub-grid physics is implemented, while in
the following three runs, termed ``Noblast'', ``Blast'' and ``Blast+'', we
included both gas cooling and star formation (with associated SNa
feedback). In these runs we assumed a density threshold for star formation
(SF) of $2\times 10^6$ cm$^{-3}$, and a typical (SF) timescale of $1.0$
Myr. We employed a SNa yield of 15\%. The specific energy budget from SNae,
$10^{50} \rm\, erg/M_\odot$, is totally released in the parent cell as thermal
energy. The subsequent SNa feedback was implemented in two different ways. The
first recipe (``Noblast'' run) assumes that gas starts to cool down
immediately after it is released in the SNa event.  The second feedback scheme
used in the ``Blast'' and ``Blast+'' runs \citep{teyssier13} assumes that the
internal energy injected by SNae is decoupled from the standard gas cooling,
decaying exponentially on a timescale of 20 Myr. This second scheme implicitly
assumes the presence of non-thermal processes accelerating the SNa blast
wave. Such non thermal energy is characterised by much longer dissipation
time-scales than the thermal component \citep[see, e.g.][]{ensslin07}.  We
finally note that typical timescale for the onset of SNae is much longer
compared to the typical gas inflow timescale in our simulations, i.e., SNae
would have little/negligible effects on MBH and gas dynamics. In order to
enhance feedback effects, we assumed no time delay for the onset of SNae after
star formation. No AGN feedback has been included in any of the runs. Our new
dynamic refinement criterion is implemented in the two ``+'' runs.

\section{Results}
 
\begin{figure}
\subfigure
{\includegraphics[width=0.238\textwidth]{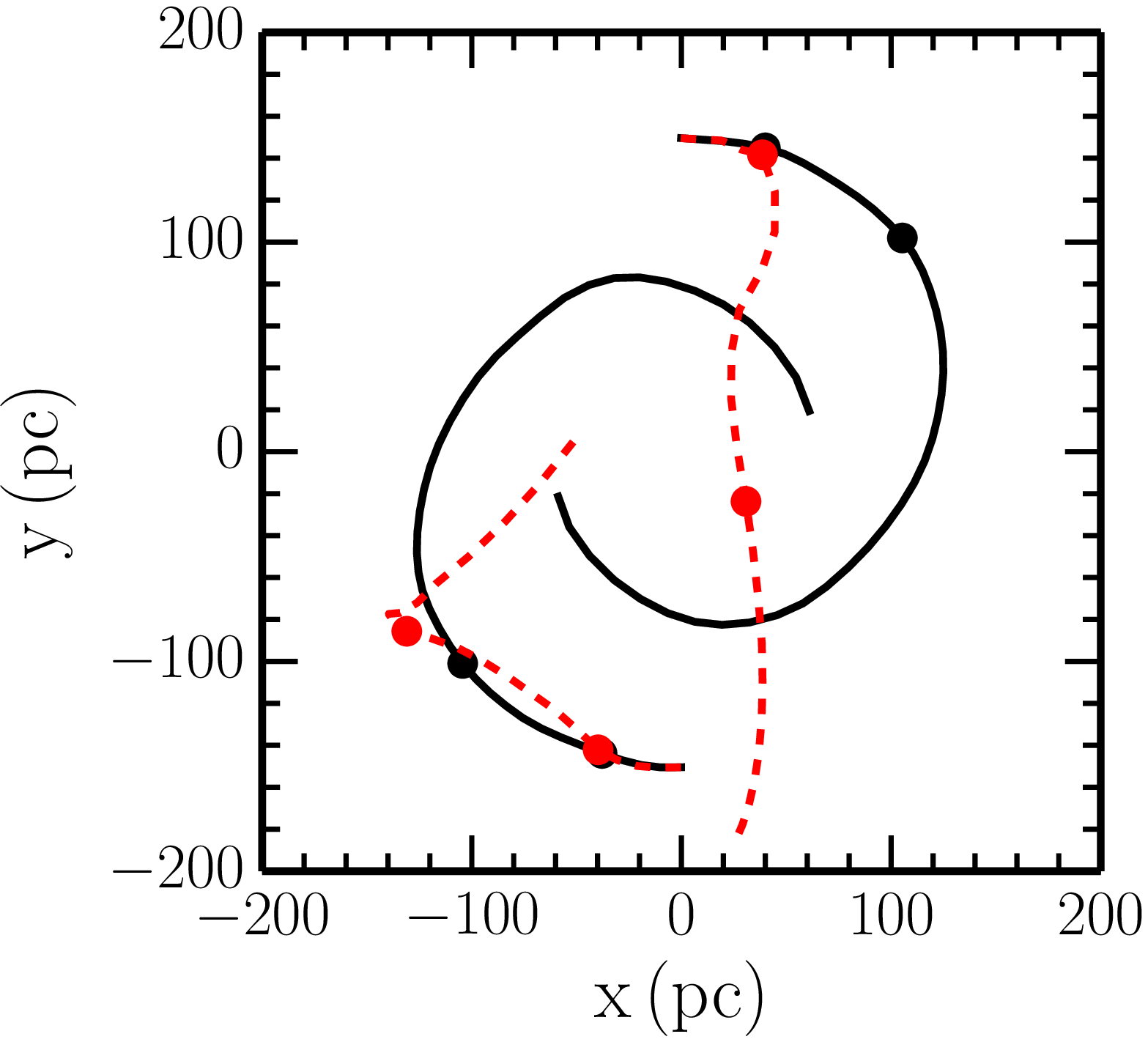}}
\subfigure
{\includegraphics[width=0.232\textwidth]{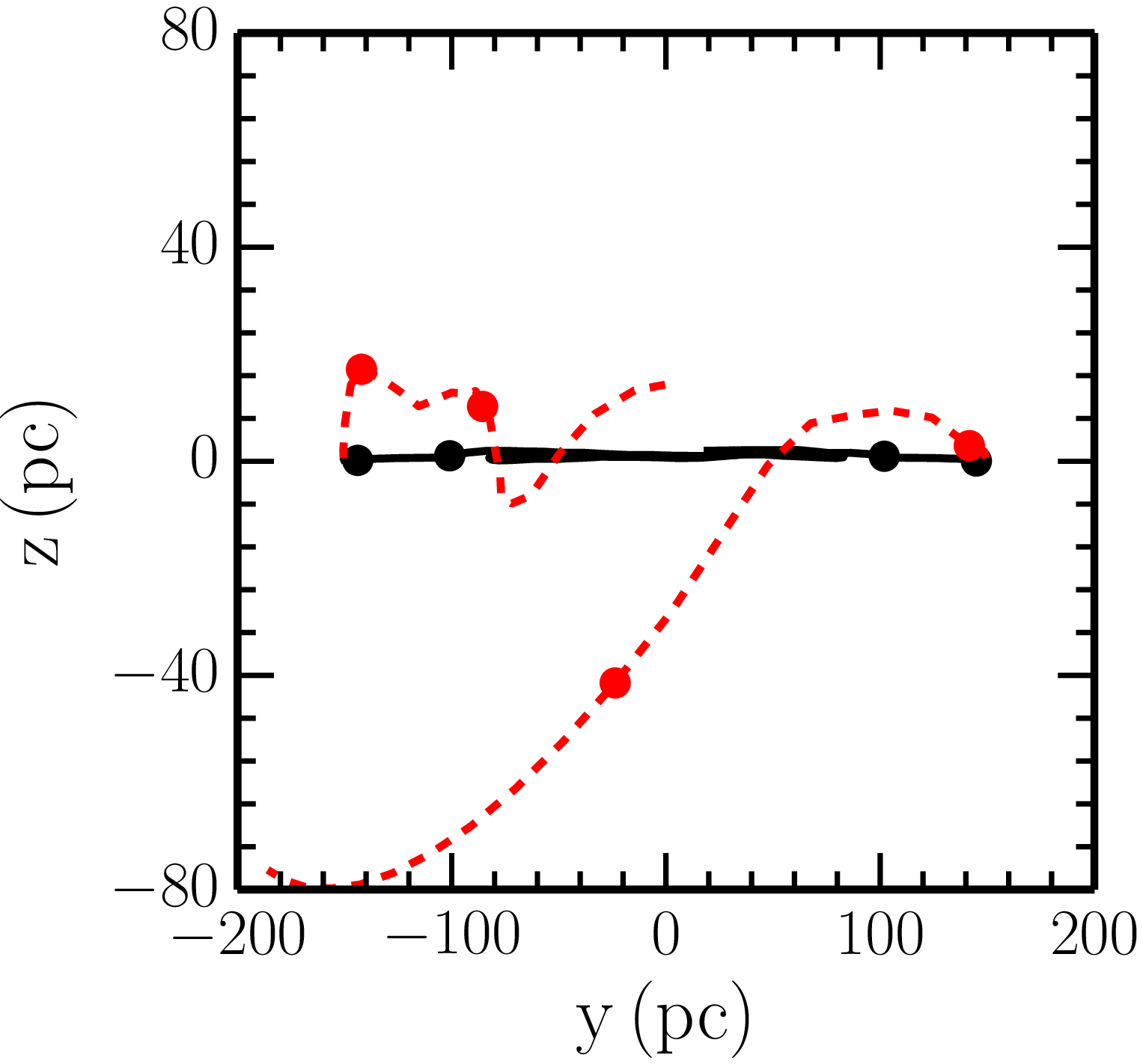}}
\subfigure
{\includegraphics[width=0.238\textwidth]{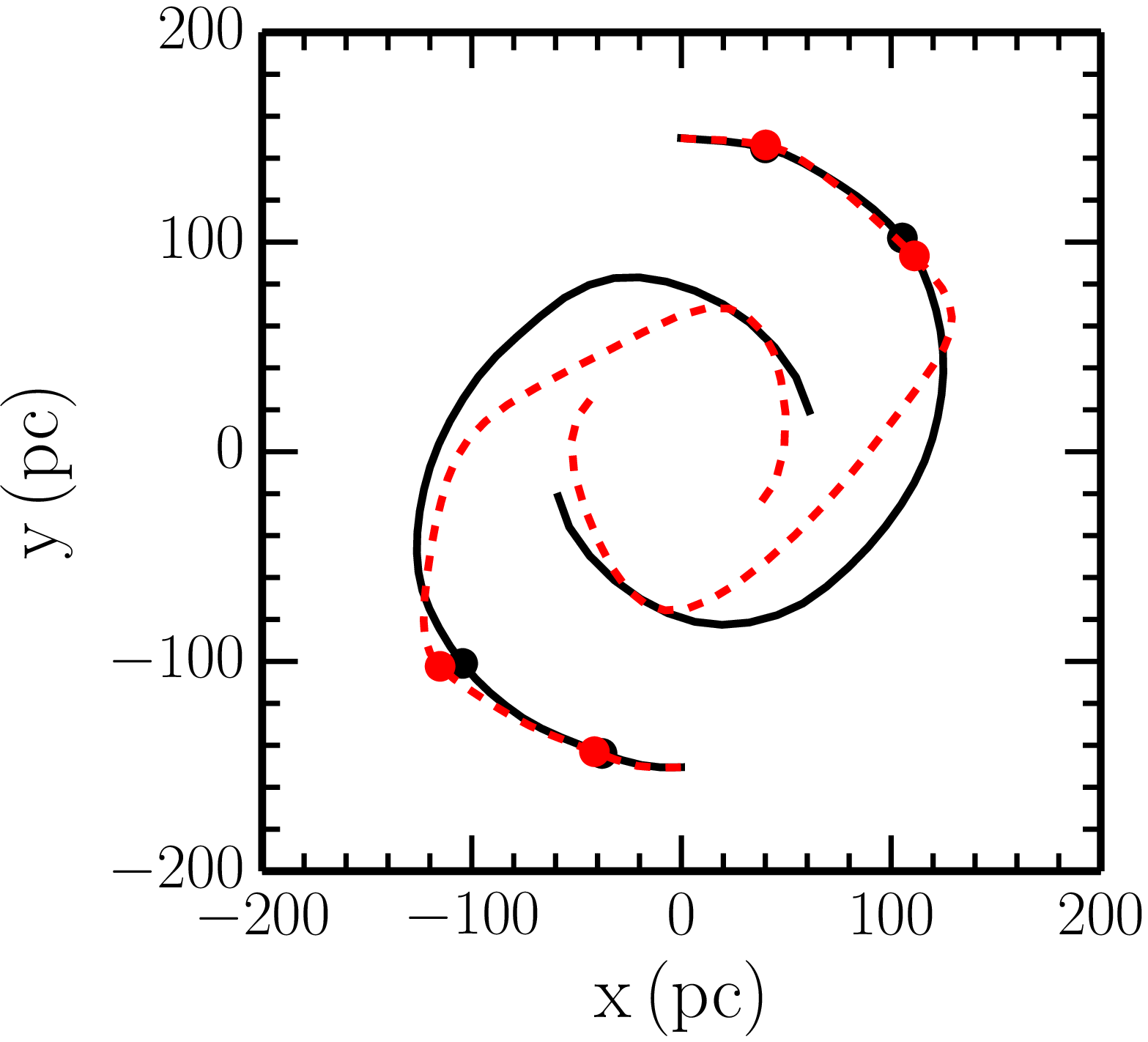}}
\subfigure
{\includegraphics[width=0.232\textwidth]{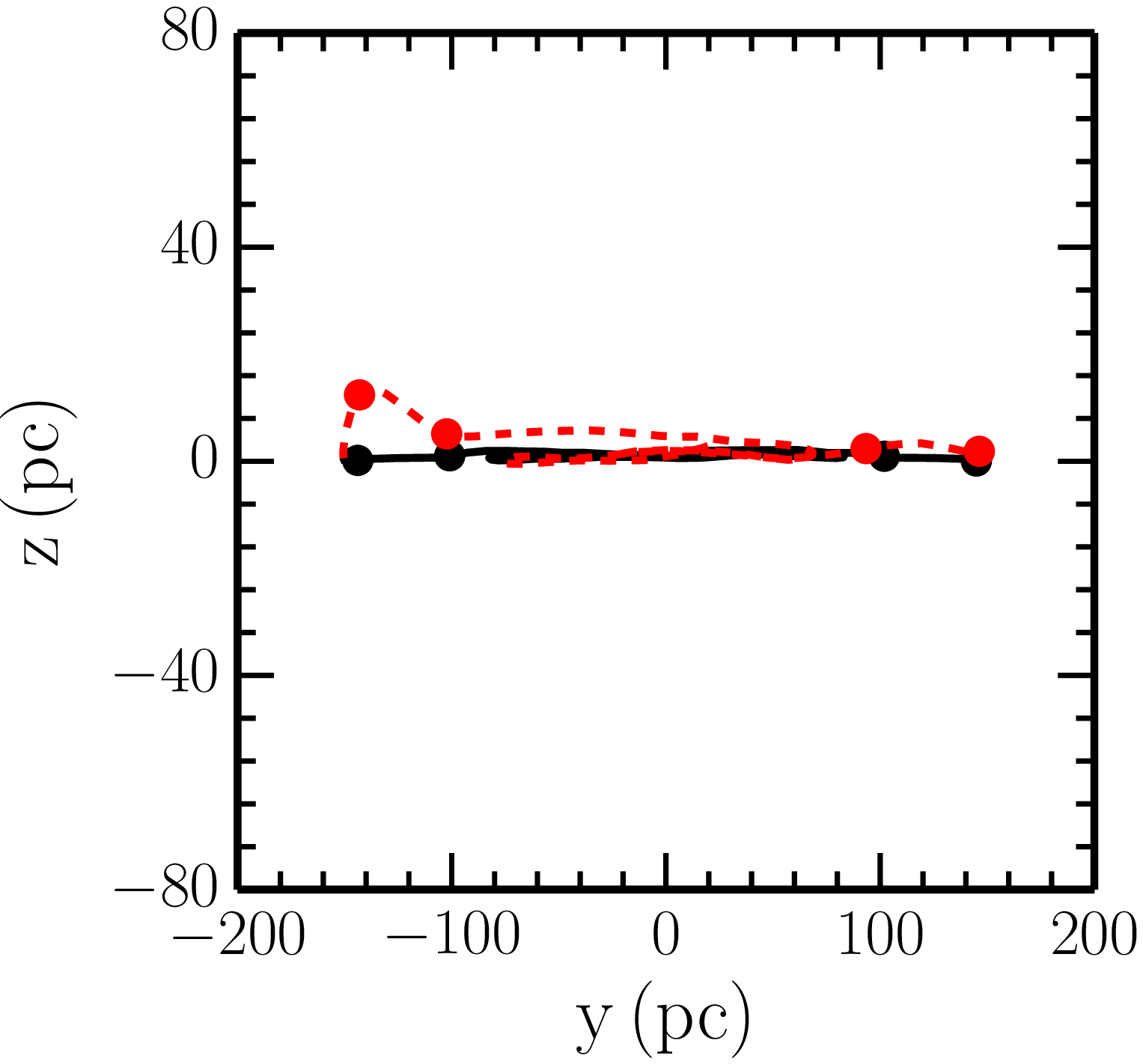}}
\caption{\small Orbits for the two MBHs from RAMSES runs Plain and Plain+ , compared
  with the SPH run. The panels on top show the orbits projected in the
  face-on (on the left) and edge-on (on the right) views for run Plain, plotted
  as red dashed lines and the SPH run, plotted as black solid lines. The
  panels at the bottom are the same plots obtained from run Plain+ and the
  SPH run. The points mark the MBH positions at $t=1$ and 3 Myr for the
  runs considered.}
\label{fig:orbitPlain}
\end{figure}

\begin{figure}
\subfigure
{\includegraphics[width=0.475\textwidth]{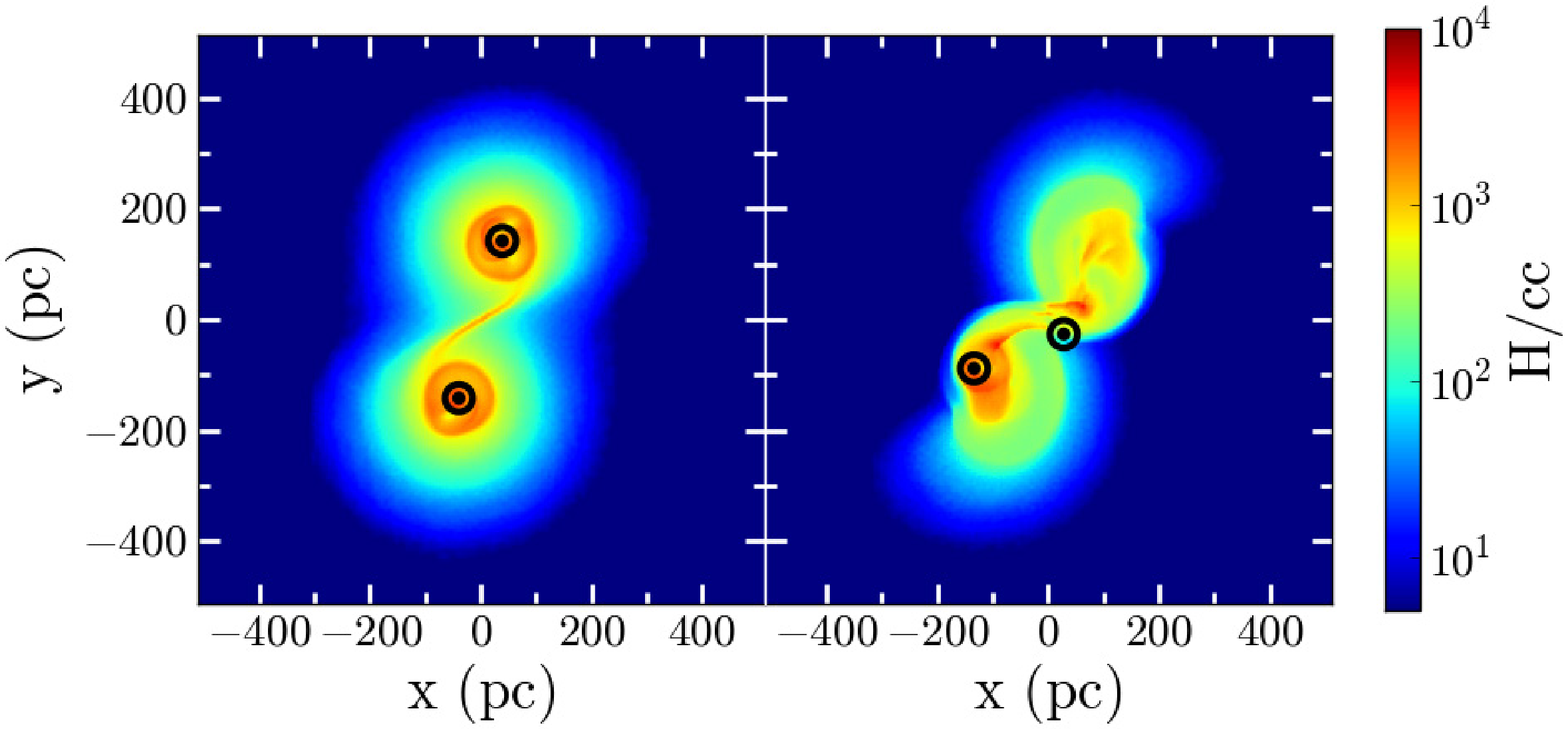}}
\subfigure
{\includegraphics[width=0.475\textwidth]{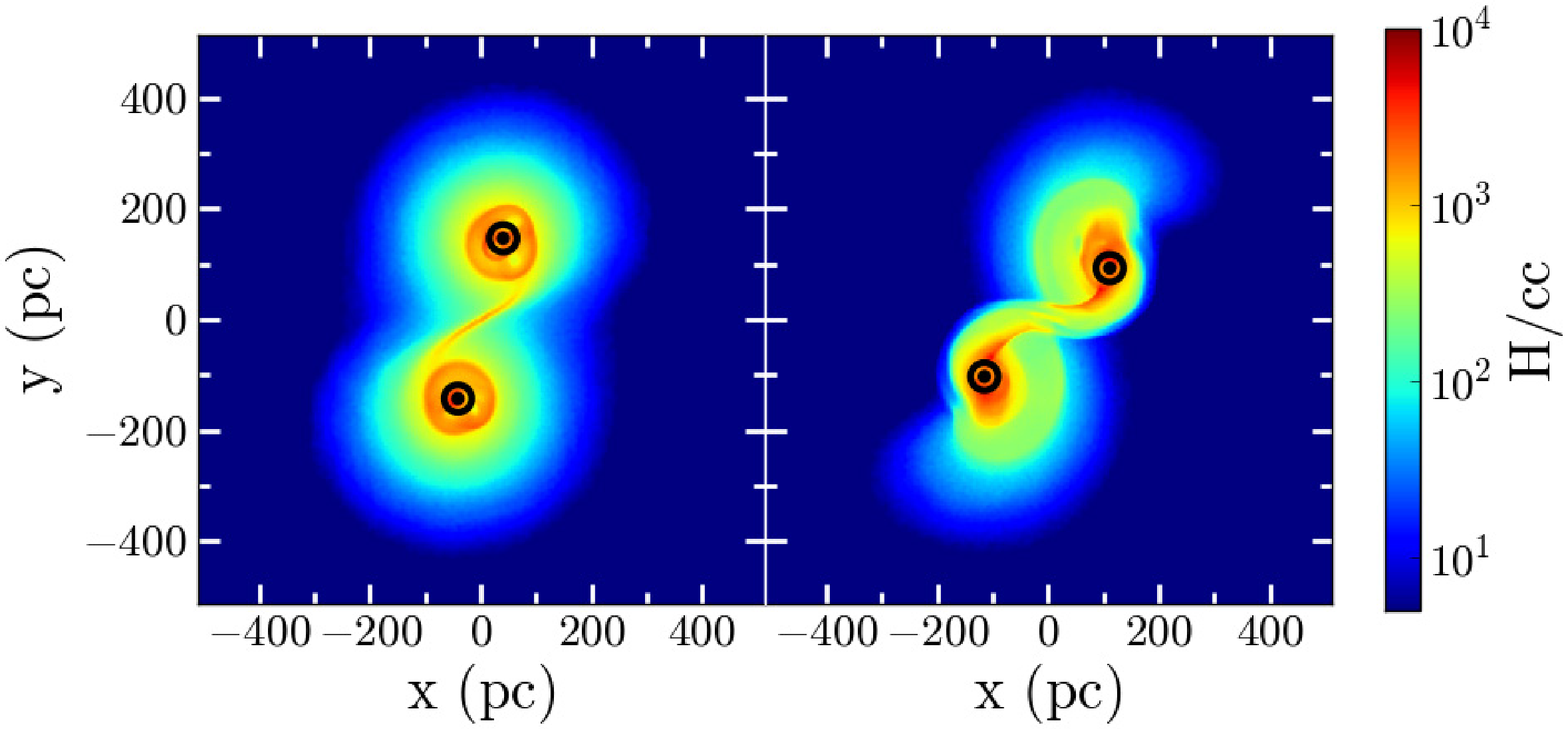}}
\caption{\small Upper panels: Gas density map at $t=1$ Myr (left panels) and $t=3$ Myr (right
  panels) for run Plain. The MBH positions are identified by the black bullseye symbols. Lower panel: same as upper panel for run Plain+.}
\label{fig:mapPlain}
\end{figure}

The upper panels of Figure~\ref{fig:orbitPlain} show the MBH pair orbit in run
Plain compared to the SPH run. While the orbital evolution computed by GADGET2
shows a smooth orbital decay of the pair, run Plain shows an abrupt change in
the direction of motion of the two MBHs after $\sim 2-3$ Myr from the
beginning of the run. At this time the MBHs suddenly leave the gas (upper
panels in Figure~\ref{fig:mapPlain}) and stellar overdensities they
inhabited. Such an abrupt acceleration could, in principle, have a physical
explanation. For example, the sudden swerve could be the outcome of short
range encounters between the MBHs and compact massive clumps or stellar
clusters. We note, however, that such an interpretation is unlikely because of
two reasons: $(i)$ a strong gravitational perturbation would have affected the
gas and stellar nuclei as well as the MBHs, and $(ii)$ as described in
Section~\ref{sec:ic}, the gaseous discs in our simulations are initially
stable against fragmentation, and the gas distribution is expected to remain
smooth during the entire evolution in run Plain, in which no cooling
prescription is implemented. A search for gas and stellar clumps in the
snapshots of run Plain confirmed this expectation.

\begin{figure*}
\includegraphics[width=\textwidth]{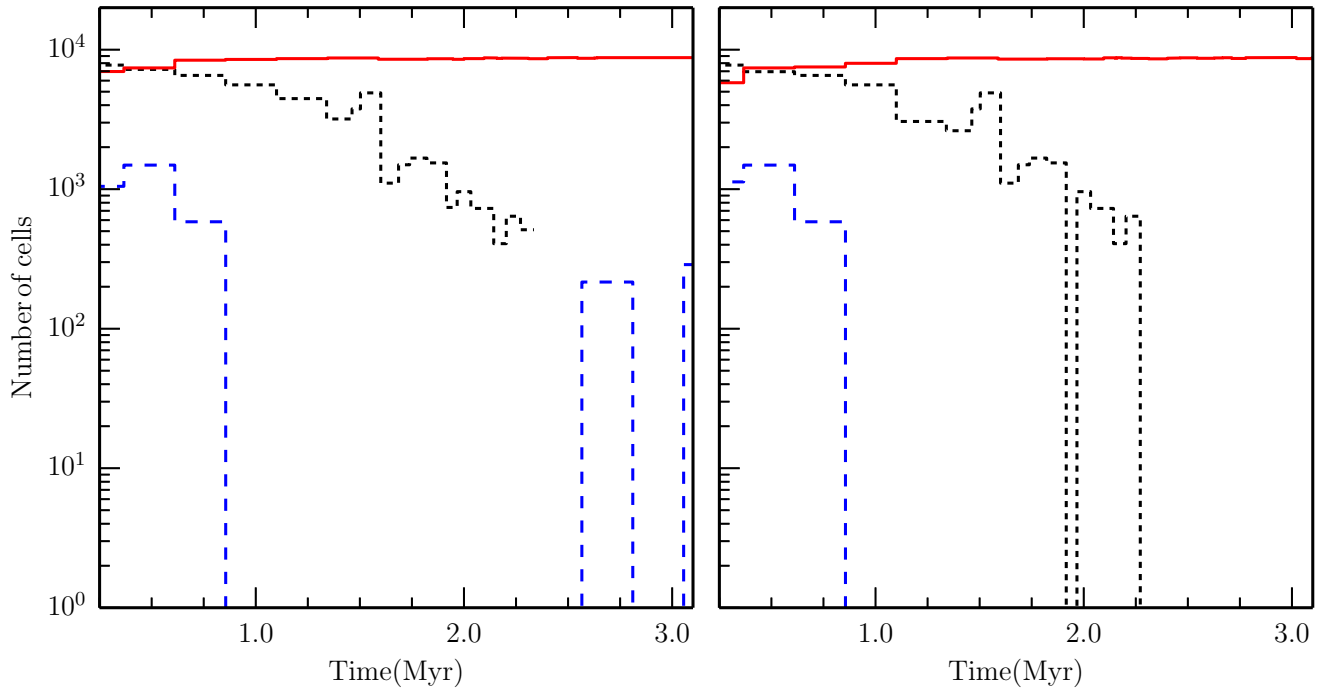}
\caption{\small Number of cells for the highest refinement level around
 each of the two MBHs versus time for simulations Plain (blue dashed line), Noblast (red solid line) and Blast (black dotted line).}
\label{fig:level}
\end{figure*}

The peculiar and unexpected dynamical evolution of the MBHs in run Plain could
be a numerical artefact, due to the rapid variation of the spatial resolution
around the two MBHs. Figure~\ref{fig:level} shows the
  number of cells at the maximum refinement level within 5 pc from each MBH. 
  The sudden drop of
  resolution is caused by a density drop during the first stages of the
  simulation. Such a gas readjustment is expected, since the initial
  conditions were stable in isolation, and the two circum--nuclear discs are
  initially set at a finite separation. We stress that, although this initial
  gas evolution is driven by the procedure used to generate the initial
  conditions, similar sudden resolution changes are expected also due to the
  evolution of the gas subject to additional physics, such as SNa explosions,
  as discussed below.

To check if the unexpected behaviour of the MBHs is a pure numerical effect we
ran the same simulation forcing the code to keep a high resolution close to
the moving MBHs, through our new refinement implementation. The MBH orbital
evolution resulting from this check (Run Plain+) is shown in
figure~\ref{fig:orbitPlain}. Run Plain+ shows a dynamical evolution closer to that
obtained in the SPH run, that by construction is not affected by any
significant fluctuation of the gravitational spatial resolution.
Figure~\ref{fig:mapPlain} shows that with our new refinement implementation the
MBHs do not decouple from the gas structure they are hosted in.  We further
stress that an enhanced resolution close to the MBHs would facilitate the
formation of gas clumps as well as maximise the effect of their gravitational
interaction (if clumps would form) with the MBHs.  The absence of abrupt kicks
in the MBH dynamics in run Plain+ proves that the MBH noisy motion observe in run
Plain is numerical and it is caused by poor/rapidly changing resolution in the
region surrounding the MBHs. Still some differences in the orbital evolution of the MBHs in run
Plain+ and SPH are observable. 
The initial difference in the vertical motion is probably caused by the resolution increase occurring in the very early stages of the simulation, when the initial conditions (with a maximum resolution of $\sim 1.5$ pc) are further refined to reach the desired resolution of $\sim 0.39$ pc.
Furthermore, the MBH orbital decay after the first 3 Myr is faster in the the
Plain+ run with respect to the SPH run. We checked that this is due to the
different magnitude of the gas inflow toward the geometrical centre of the
system. Such inflows are caused by the angular momentum removal associated
with the shocks developing at the contact surface between the two merging
circum--nuclear discs (CNDs).  The two numerical implementations (SPH and AMR) differ significantly in their treatment of the shocks, resulting in a different MBH dynamics. The detailed discussion of the physical evolution of the system and of its effect on the pairing of the MBHs will be discussed in Lupi et al. in prep.
 

\begin{figure}
\subfigure
{\includegraphics[width=0.238\textwidth]{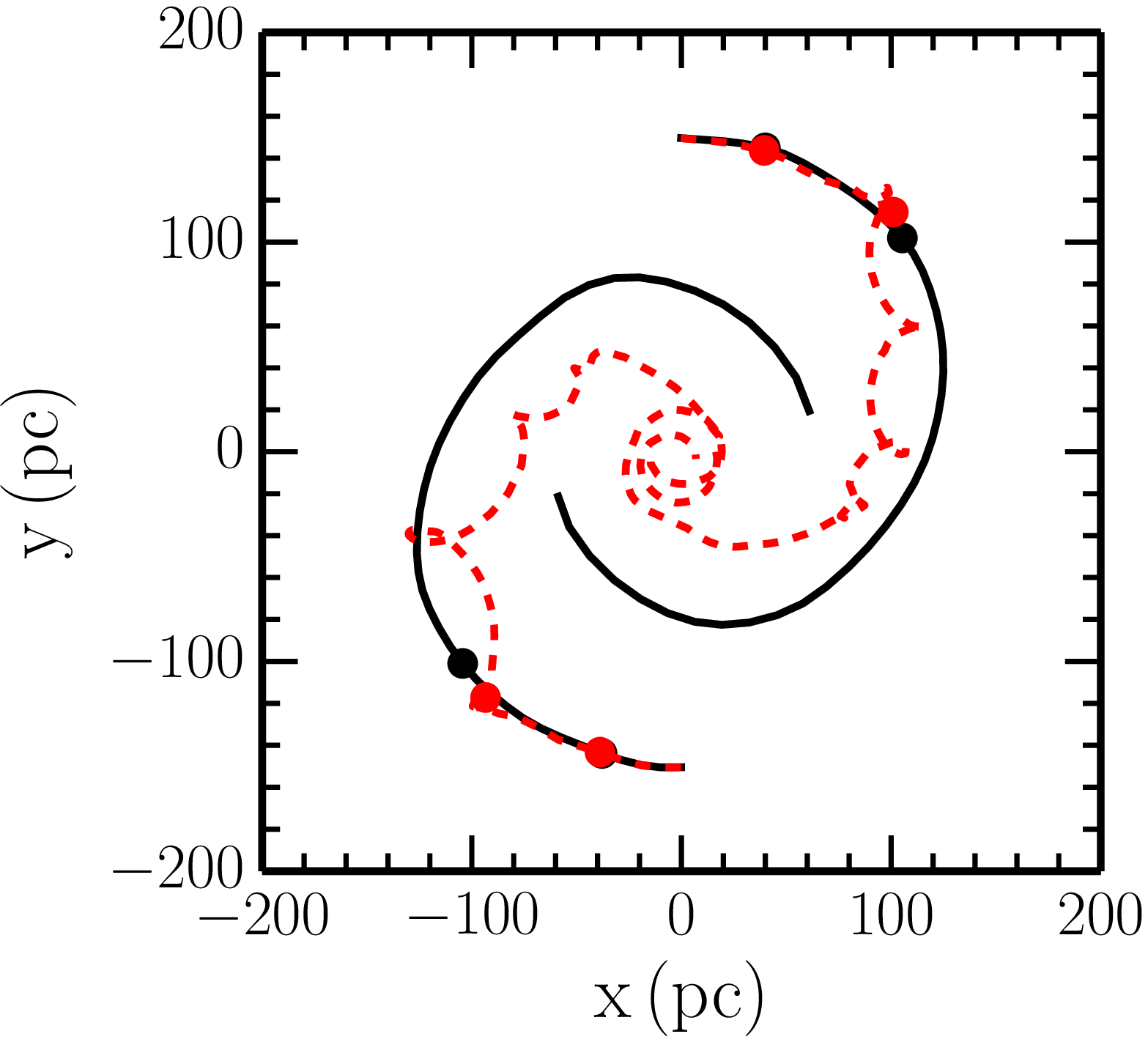}}
\subfigure
{\includegraphics[width=0.232\textwidth]{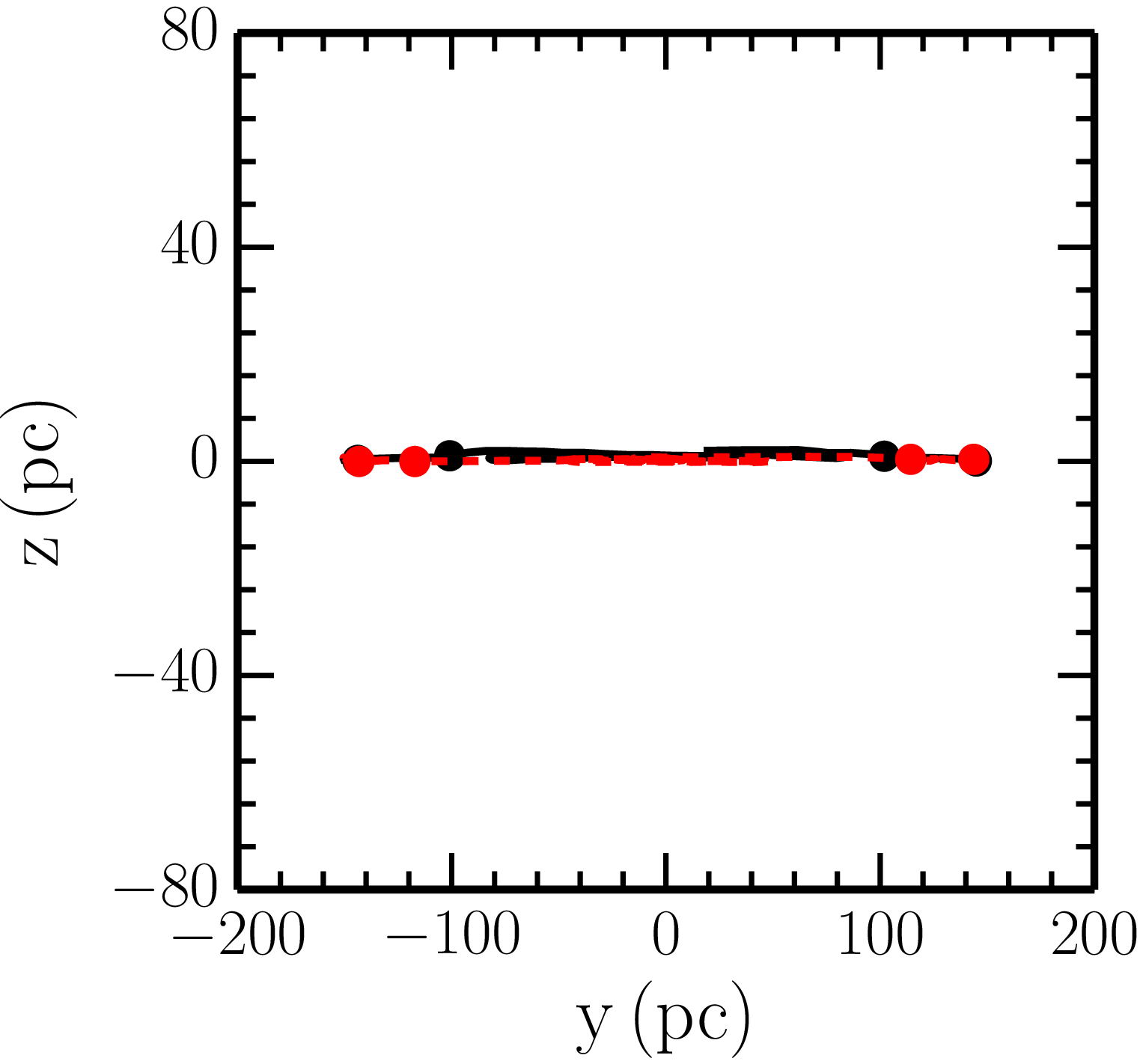}}
\caption{\small 
Same as figure~\ref{fig:orbitPlain} for the Noblast/SPH runs comparison. 
}
\label{fig:orbitNoblast}
\end{figure}

\begin{figure*}
\subfigure
{\includegraphics[width=1.0\textwidth]{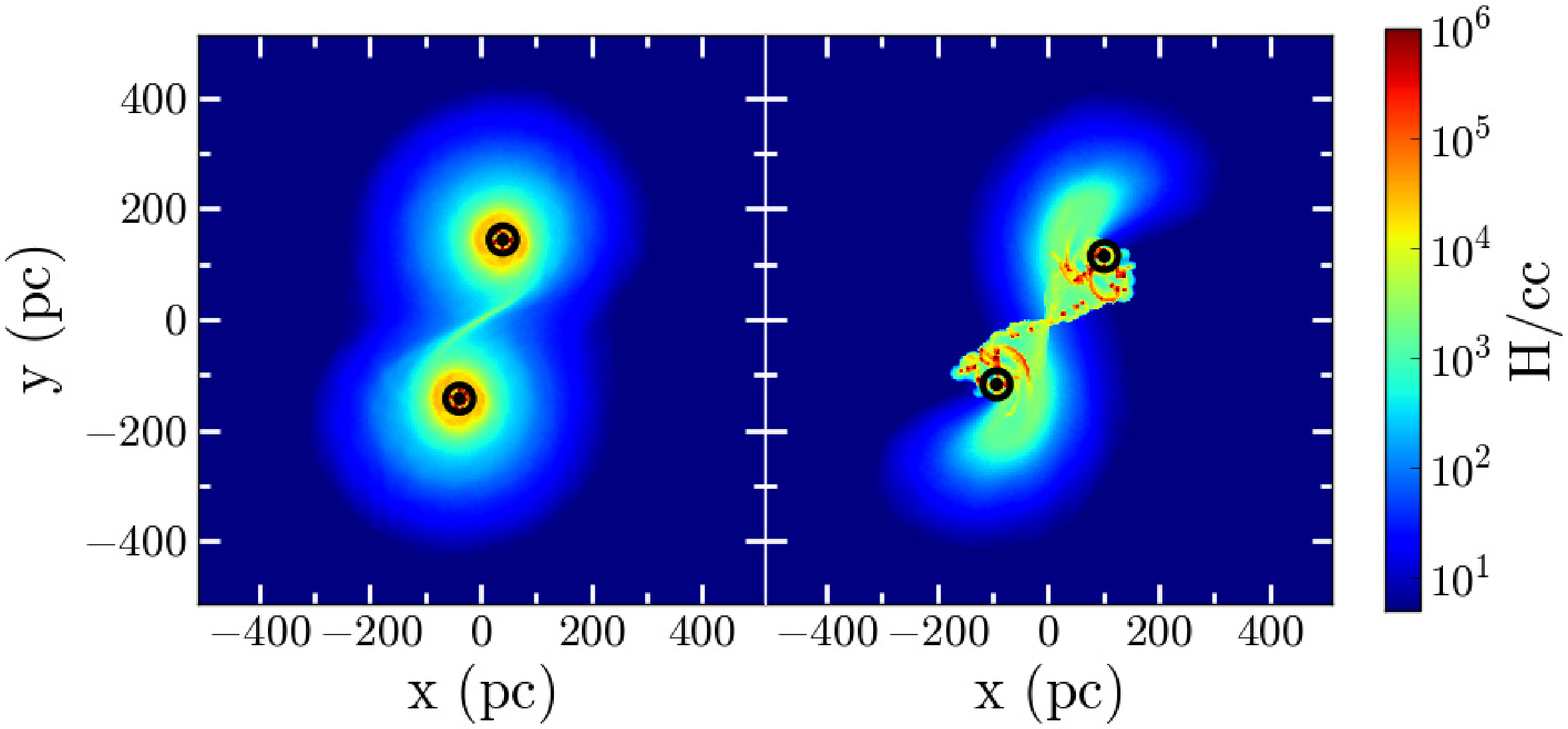}}
\subfigure
{\includegraphics[width=1.0\textwidth]{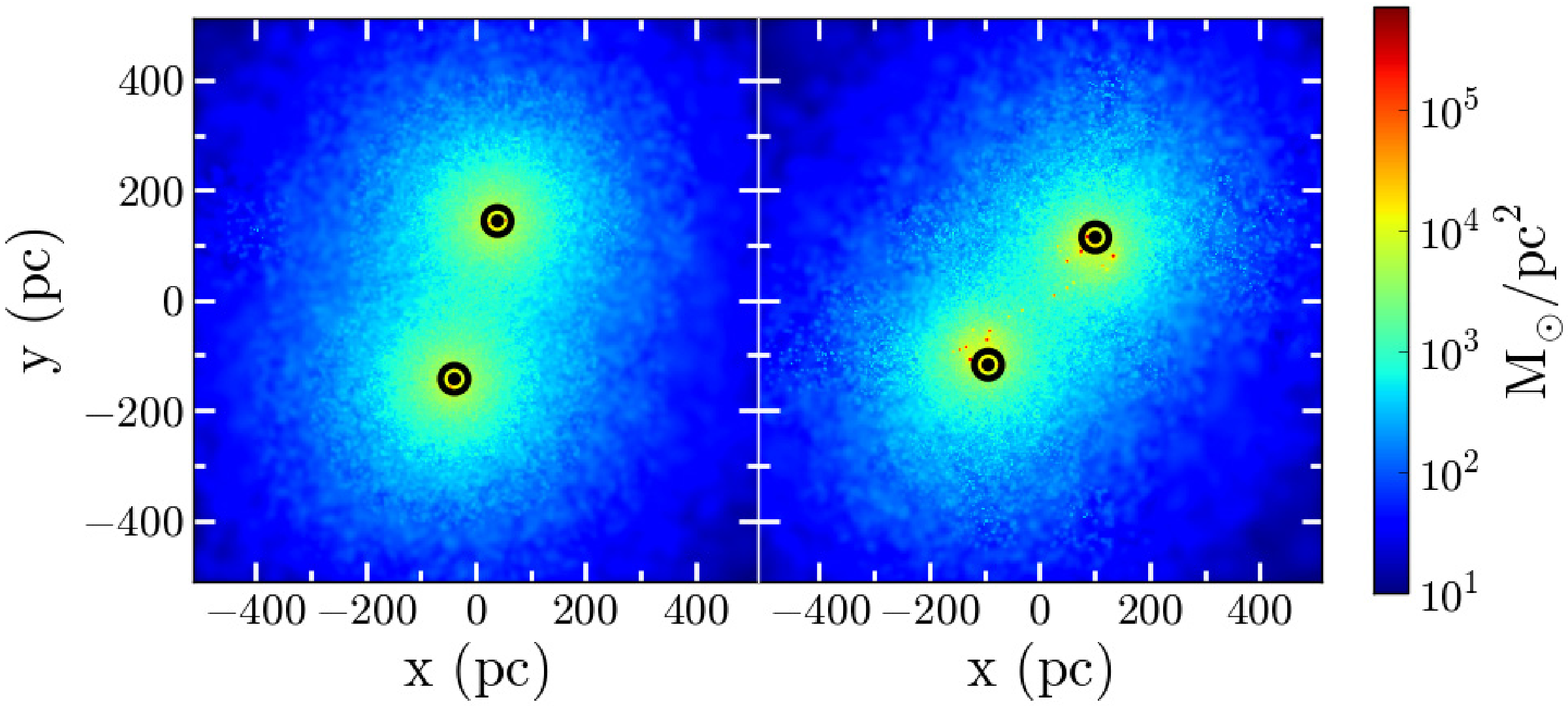}}
\caption{\small  Upper (lower) panel: gas (stellar) density map at $t=1$ and 3 Myr (left and right panel
  respectively) in run Noblast. 
  All the notation is the same as in figure~\ref{fig:mapPlain}}
\label{fig:mapNoblast}
\end{figure*}

To study the effect of the refinement prescriptions onto the MBH dynamics in
less idealised simulations, we performed three runs (Noblast, Blast and
Blast+) allowing the gas to radiatively cool and form stars. As shown in the
following, the orbital evolution strongly depends on the different
implementation for the feedback by SNae. Figure~\ref{fig:orbitNoblast} shows the MBH
orbital evolution in run Noblast. The MBH dynamics does not show anything
similar to the huge kicks that decouple the MBH dynamics from the gas
distribution observed in run Plain. On the contrary, figure~\ref{fig:mapNoblast}
demonstrates that the MBHs are still well within the gas and stellar overdensities close to
the centres of the dramatically perturbed nuclear discs. 

However, smaller swerves mainly limited to the disc plane are still observable
in the MBH orbits (figure~\ref{fig:orbitNoblast}). Figure~\ref{fig:level}
demonstrates that, in run Noblast, the wiggles in the orbits are
not related to a decrease of the spatial resolution. Indeed the resolution
around each MBH remains almost constant during the entire run, with a high
number of cells populating the maximum refinement level. Such a high
resolution is ensured by the formation of high density condensation of cooling
gas around the MBHs.

\begin{figure*}
\includegraphics[width=\textwidth]{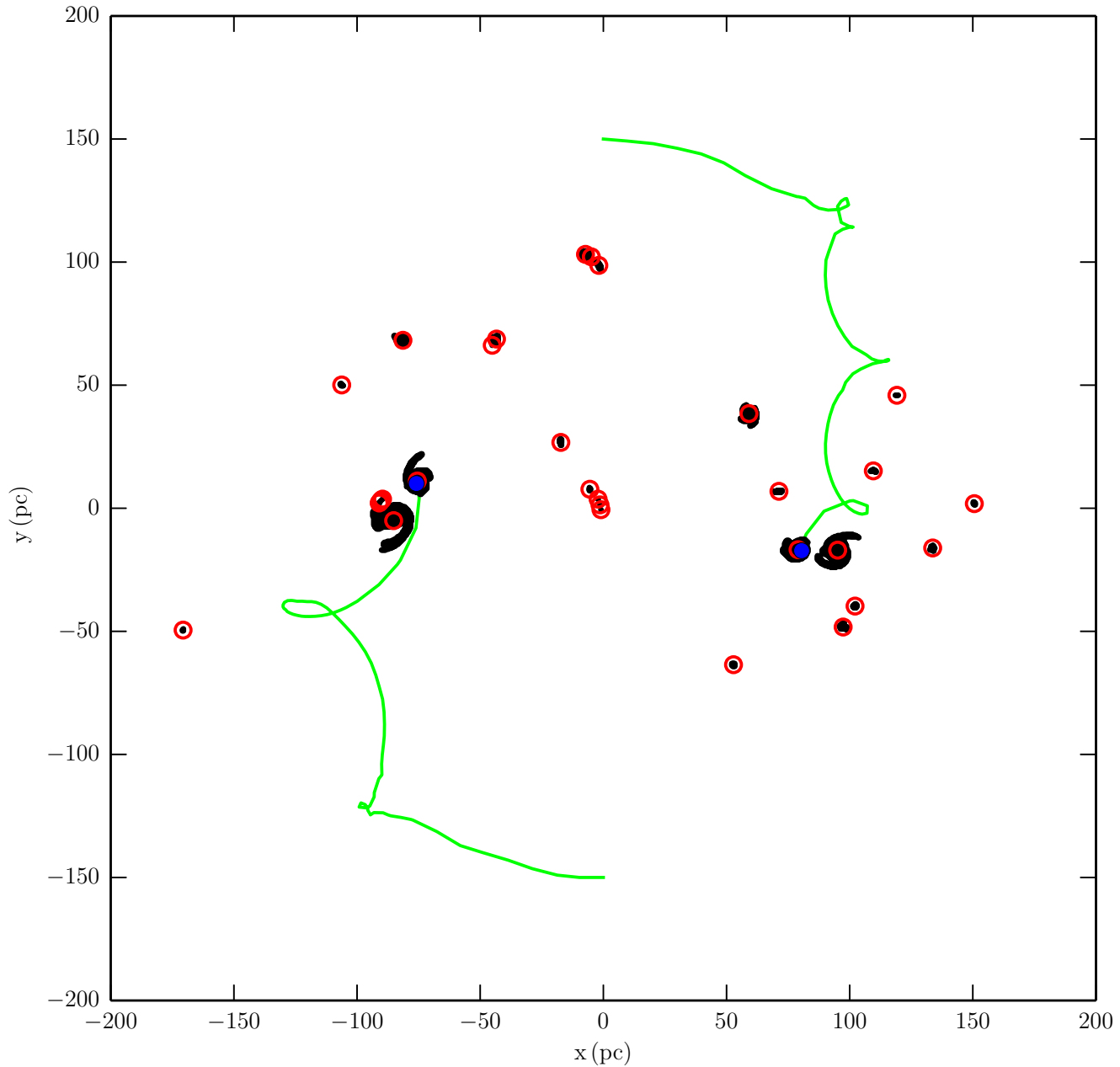}
\caption{\small  Strong interactions between the MBHs and massive gas clouds in run Noblast at time 5.8 Myr. The MBHs orbital path and current positions are marked with green lines and blue dots. The black regions highlight the cells forming the clouds, whose centre of mass is marked by the red empty circles (only for clouds formed by at least 10 cells).}
\label{fig:clouds1}
\end{figure*}

\begin{figure*}
\includegraphics[width=\textwidth]{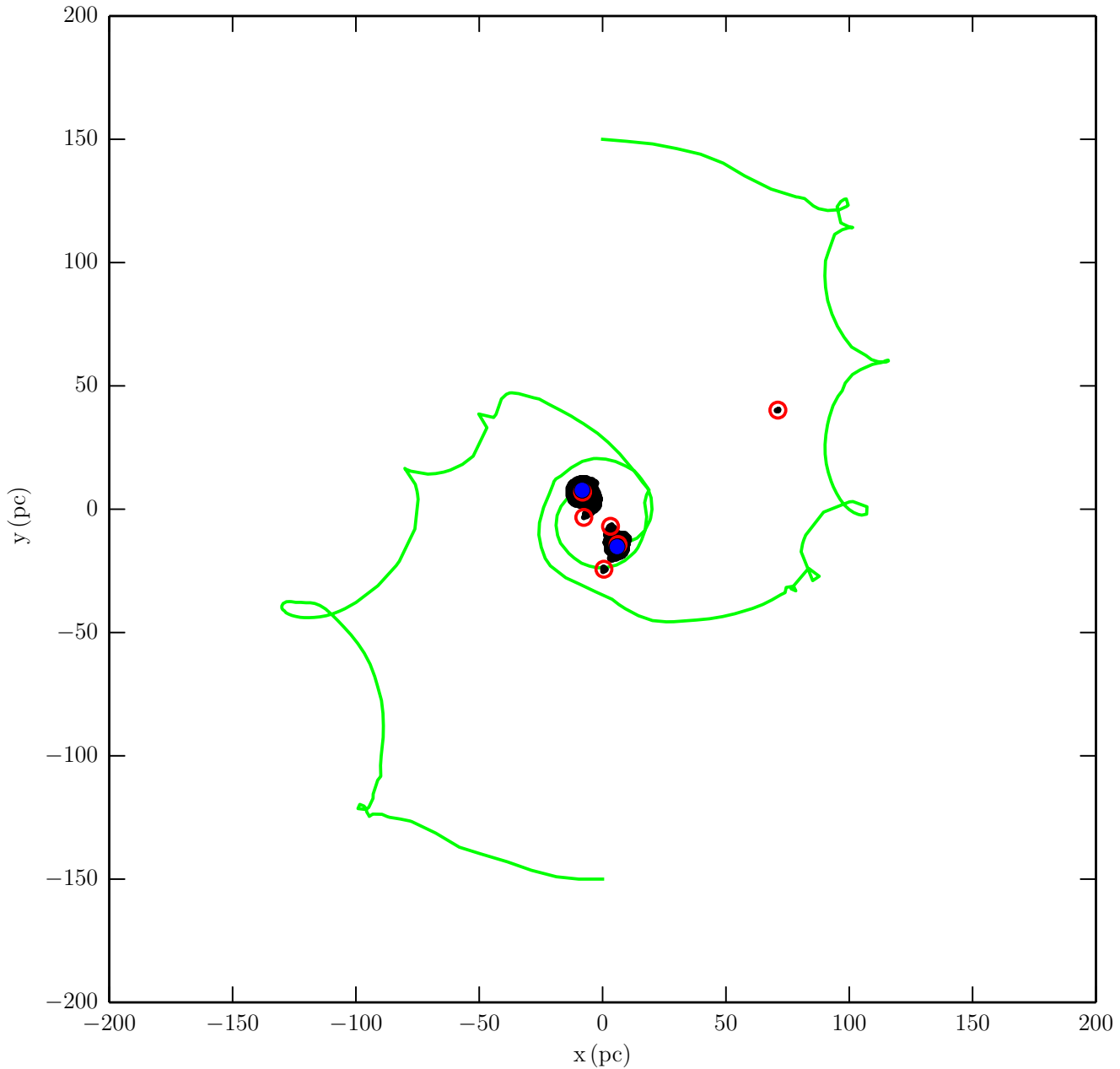}
\caption{\small Same as figure~\ref{fig:clouds1}, but at time $t=9$ Myr. The figure shows a final stage of the orbital evolution, when the MBHs are surrounded by gas overdensities and no significant  MBH/cloud interactions are taking place.}
\label{fig:clouds2}  
\end{figure*}

The peculiarities in the MBH orbits in run Noblast are due to close
interactions with massive clumps, forming in the disc when the gas is allowed
to cool. Indeed a large number of massive clumps form during the first stages
of the merger, especially along the gas shock surface between the two gaseous
discs, as observable in the left panel of figure~\ref{fig:mapNoblast}. These clumps
can lead to very energetic kicks to MBHs, unless they are destroyed before the
interaction by SNae.  This is not the case for run Noblast, in which the large
thermal energy injected by SNae in the gas immediately starts to cool, thus
leaving clumps nearly unperturbed. As discussed above, this boosts the
probability of having a strong cloud/MBH interaction, and results in a high
resolution close to the MBHs (preventing spurious numerical wandering of the
MBHs).  Figure~\ref{fig:clouds1} highlights a strong interaction between each MBH and a massive cloud in run Noblast, taking place at $t=5.8$ Myr. Figure~\ref{fig:clouds2} shows a later stage ($t=9$ Myr) of the evolution, when the two MBHs evolve in a smoother environment. In both cases the clouds have been identified extracting the cells with a density exceeding $8\times 10^5$ H/cc and then grouping together the adjacent cells.
The detailed analysis of the interactions between MBHs and clouds as well as a broader study of the effect of the gas dynamics onto the MBH pairing is deferred to a paper in preparation.

\begin{figure}
\subfigure
{\includegraphics[width=0.238\textwidth]{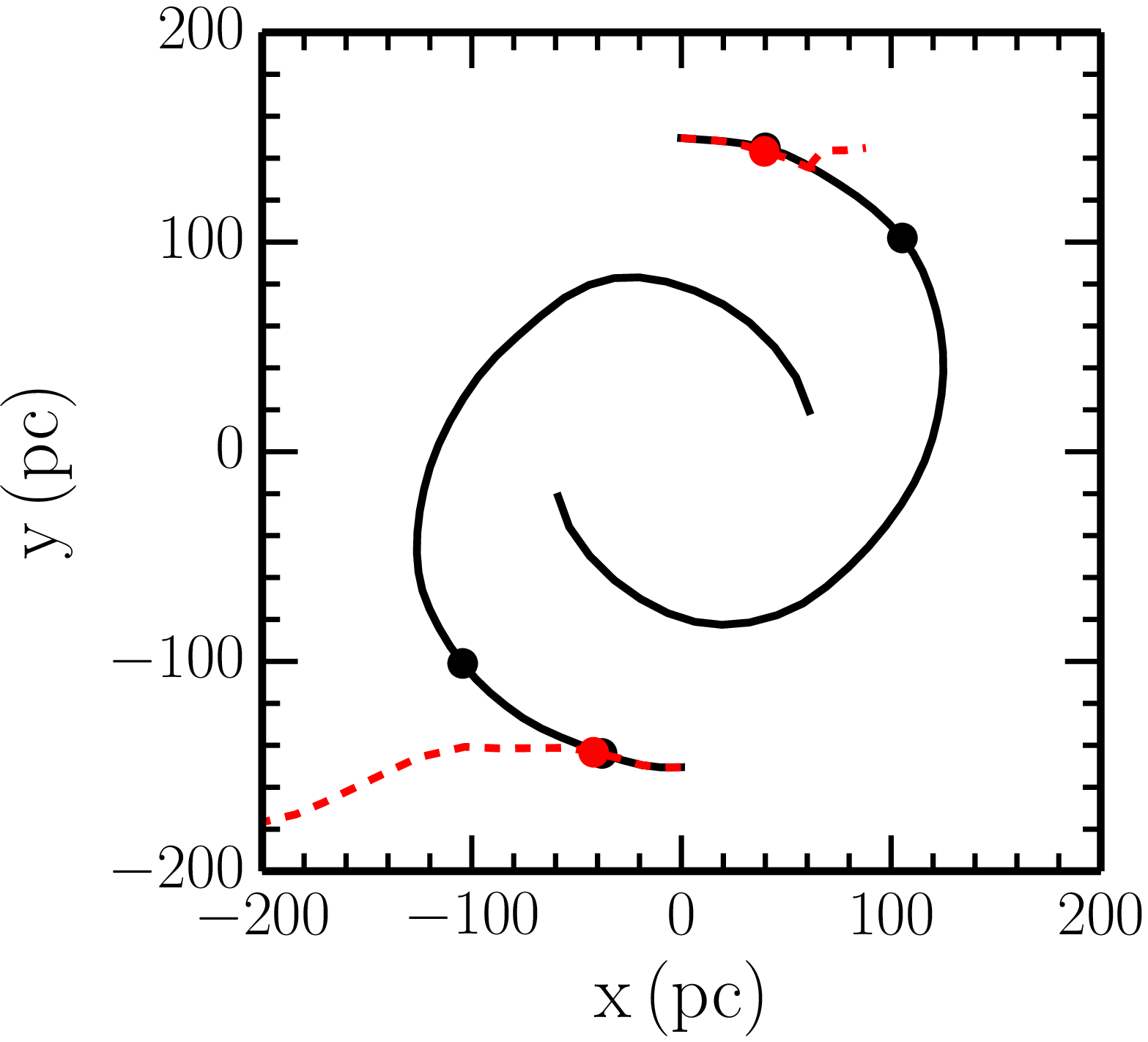}}
\subfigure
{\includegraphics[width=0.232\textwidth]{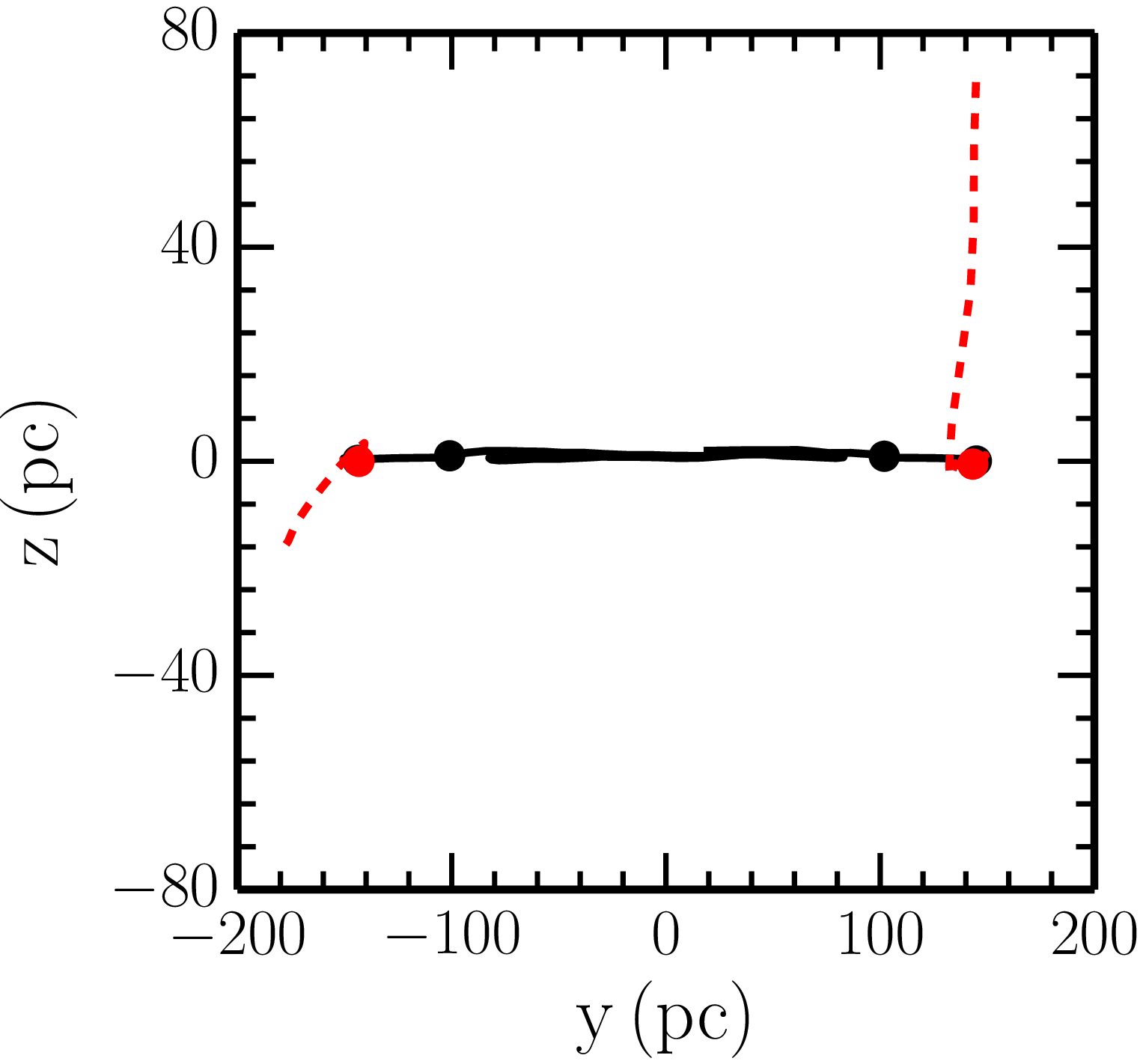}}
\subfigure
{\includegraphics[width=0.238\textwidth]{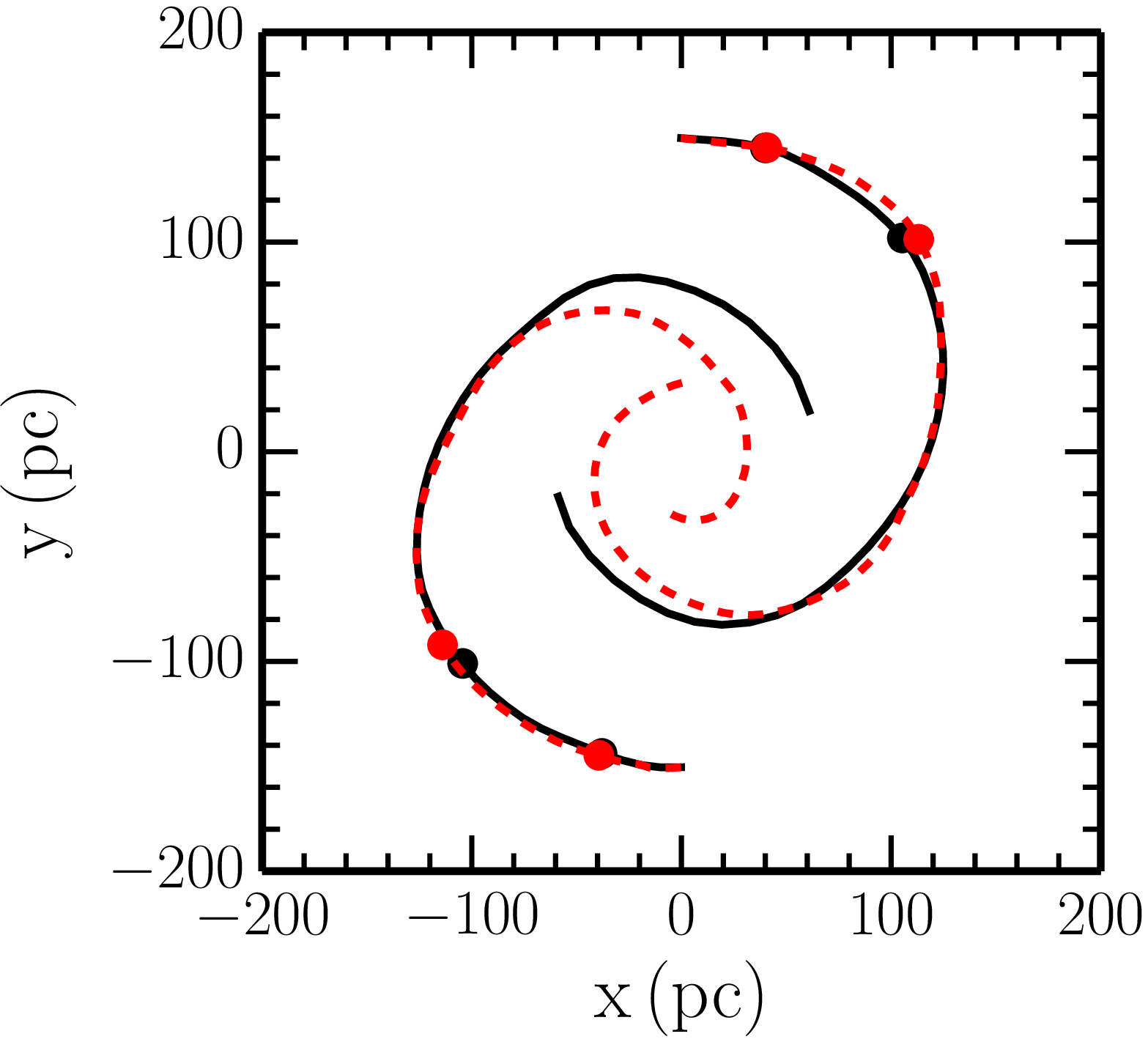}}
\subfigure
{\includegraphics[width=0.232\textwidth]{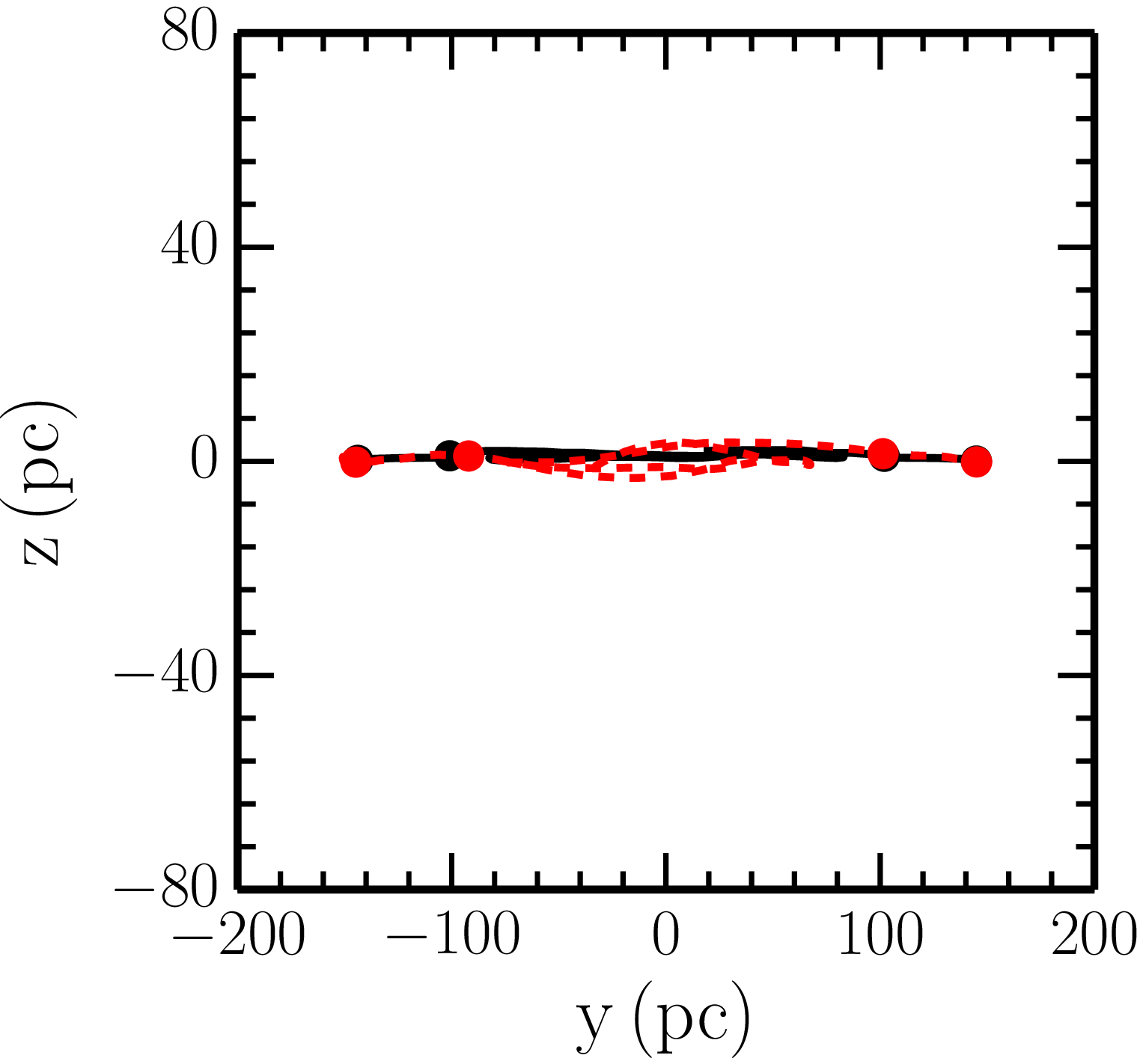}}
\caption{\small Same as figure~\ref{fig:orbitPlain} for the comparison between run SPH and runs Blast and Blast+.}
\label{fig:orbitBlast}
\end{figure}

If instead the gas is unable to rapidly get rid of the energy injected by
SNae we expect a smaller incidence of MBH-cloud
interactions, but at the same time the SNae can strongly affect the densest
and intensely star-forming regions close to the MBHs. A SNa driven gas
depletion may result in a decreasing force resolution when the new refinement
discussed here is not implemented. Figure~\ref{fig:orbitBlast} shows a
comparison between the MBH dynamical evolution observed in runs Blast and
Blast+. A peculiar wandering of the two MBHs in the three dimensions is
observed in run Blast, similarly to what happens in run Plain.  We stopped the
run after 2 Myr only, when the MBH motion had already been affected by the
numerical effect and MBHs had been scattered very far from the disc plane.

\begin{figure}
\subfigure
{\includegraphics[width=0.475\textwidth]{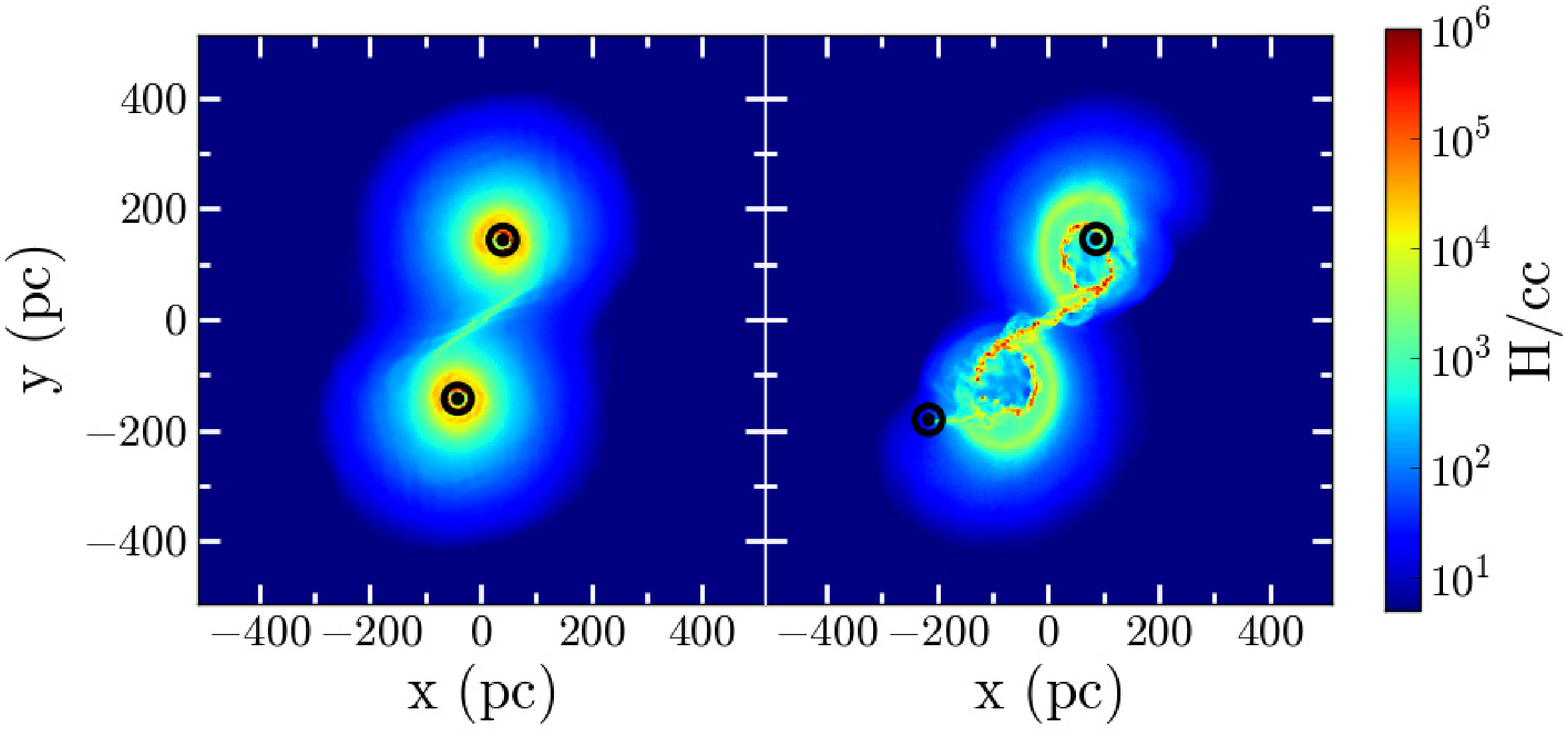}}
\subfigure
{\includegraphics[width=0.475\textwidth]{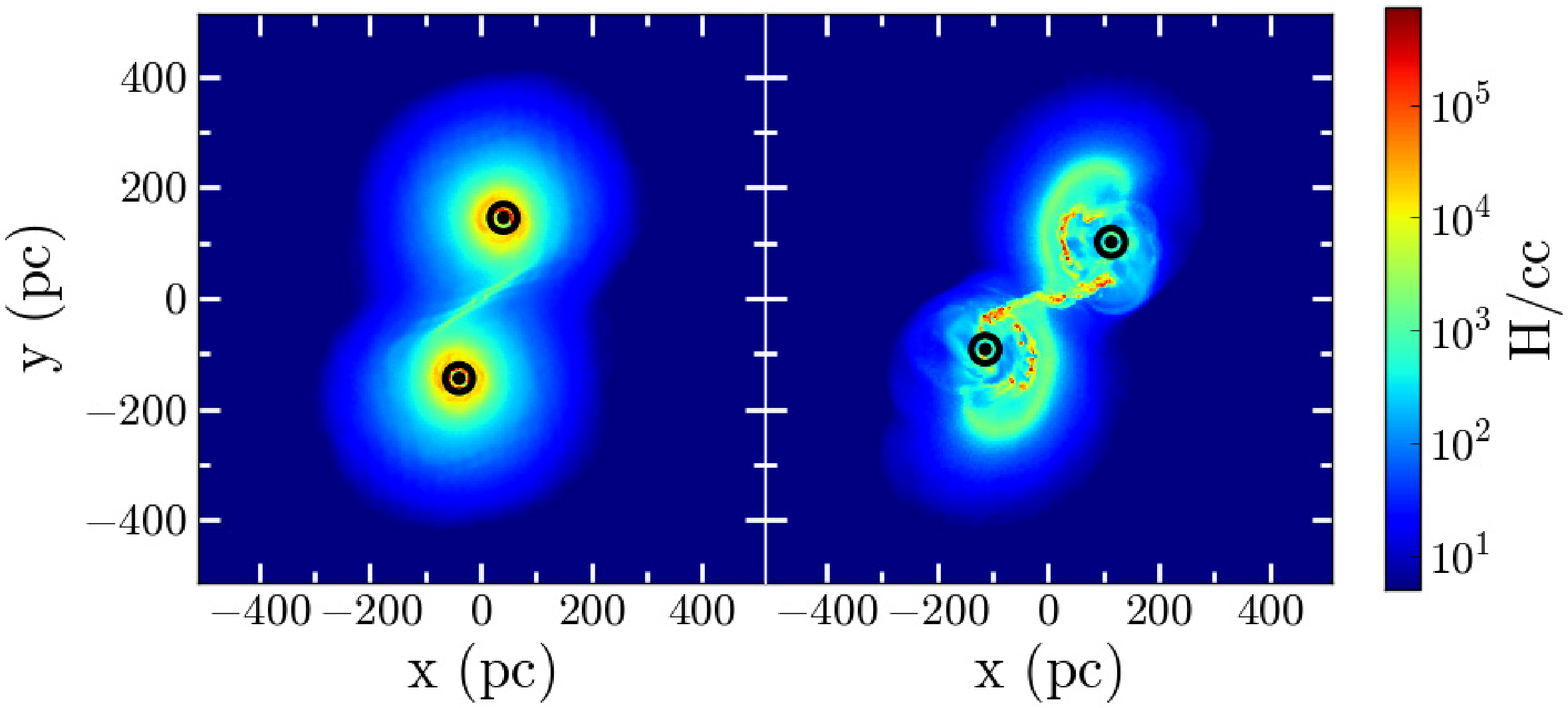}}
\caption{\small Same as figure~\ref{fig:mapPlain} for runs Blast and Blast+.}
\label{fig:mapBlast}
\end{figure}

Again, the peculiar motion of MBHs in run Blast could either be a numerical
artefact or have a physical origin. We note that in both runs Blast and Blast+
the clumps are disrupted on short timescales by SNae. Hence, gas overdensities
are not expected to perturb significantly the dynamical evolution of
MBHs. Furthermore, the feedback are energetic enough to deplete the gas from
the nuclear regions of both discs, leaving the MBHs in an under-dense region
(see figure~\ref{fig:mapBlast}). The time evolution of the number of cells at
high resolution levels in the MBH vicinities is shown in Figure
\ref{fig:level}.  This confirms that the energy injection from SNae drives a
significant resolution drop during the first 3 Myr, as also observed in run
Plain. In run Blast, however, the loss of resolution does not directly depends
on our realisation of the initial conditions, but it is a consequence of the
physical evolution of the system.

The different dynamical evolution observed in run Blast+ (lower panels in
figure~\ref{fig:orbitBlast}) finally proves that the jerks in the MBHs paths are
numerical artefacts. In facts, in this last case, the MBHs follows a very
smooth evolution over multiple orbital timescales, due to the little effect of
the transient gas overdensities onto the MBHs. The comparison between the
results of run Blast and run Blast+ proves the effectiveness of refinement
implementation discussed here in modelling massive particle dynamics in
rapidly evolving backgrounds.


\section{Conclusions}
In this paper we propose a new refinement criterion to follow in
details the motion of massive particles in AMR simulations with
RAMSES. The goal of our investigation is to achieve an accurate
dynamical evolution of MBHs under the influence of the evolving
gaseous and stellar background, as expected in the final stages of a
galaxy merger.

The accuracy in the computation of gravity force in AMR codes strictly
depends on the local resolution and thus on changes in the physical
properties of gas and particles during the runs. The refinement
prescriptions already implemented in RAMSES do not ensure a proper
description of the orbit of massive particles. Whenever the region
close to the MBHs is de-refined the MBHs can experience strong
spurious perturbations and then follow unphysical orbits, often leading out of the host nucleus.

We simulated the merger of two CND/stellar nucleus structures, each hosting a MBH.  Such system should be considered as an idealised model of the latest stages of a galaxy merger, as we neglected possible galactic scale effects (e.g., large scale gas inflows). We ran a suite of simulations, including different sub-grid physics
prescriptions for the gas, in order to test the changes in the MBH
orbits with or without our additional refinement criterion. We demonstrated that
whenever the gas density near the MBHs is reduced, the simulations ran
without our new implementation result in unphysical orbits. Such MBH
wandering was observed in runs without gas cooling as well as in the
more complex case in which the gas in the two nuclei is allowed to
cool, form stars and be efficiently heated by stellar feedbacks. We
stress that the complex, violent and intrinsically dynamical nature of
the merger does not allow to predict whether and when a sudden drop in
resolution will occur. In order to trust the simulation results a
different refinement criterion, forcing a high and constant resolution
near the MBHs, is required. The prescription we designed enforces the
region around each massive particle to remain at the maximum
resolution allowed. Such region follows the MBHs along their orbits,
reducing the computational cost of the runs, and avoiding the spurious
effect caused by the resolution changes.

Other possible solutions to the numerical noise of massive particle
dynamics have been proposed in literature \citep{gabor13,dubois14a}. We
stress that our implementation has the ability of not altering the
physical evolution of the MBHs\footnote{During the final editing of this paper a new version of RAMSES has been released. In the new implementation the accelerations experienced by massive collisionless particles are computed through direct summation (R. Teyssier, private communication). The comparison between our refinement strategy and the new implementation 
 is postponed to a future investigation.}. As an example, we demonstrated that in our
implementation any close interaction between the MBHs and
gas clumps is properly modelled (at an equivalent/better
resolution than what would be achieved without our refinement
criterion). Hence, while our prescription removes the numerical
effects resulting in an artificially noisy orbital evolution, it
preserves all the non trivial physical processes that take place in
violent dynamical scenario such as the final stages of a galaxy
merger. 
 
We plan to use the new refinement criterion in a larger suite of
simulations to accurately study the MBHs dynamics down to the
formation of a close binary, to constrain the ability of the MBHs to
accrete during their final pairing, and to properly trace the dynamics
of the fuelling gas. This is important since, e.g., accreting gas affects the MBH
spins and in case of mergers, the spin of the MBH remnant and its recoil velocity 
\citep[see e.g.][and references therein]{Sesana14}. Our
study will complement the AMR runs discussed in, e.g.,
\cite{dubois14a,dubois14b}, constraining the effect of the MBH
numerical wandering on its feeding and its spin
evolution. Furthermore, the new simulation suite will serve as a
comparison to similar investigations performed with SPH simulations
\citep{escala05, dotti09, dotti10, hopkins12, maio13}, that do not
suffer the spurious MBH dynamical evolution (since the gravitational
resolution is fixed in time) but that have a completely different
implementation of the gas hydrodynamics.

Finally, our suite will also test whether (or under which conditions)
the strong perturbations due to the merger of the two host nuclei can
result in the condensation of huge and compact gas overdensities,
possible progenitors of MBHs \citep{mayer10, ferrara13, bonoli14}.

\section{Acknowledgements}
We thank R. Teyssier and L. Paredi for many fruitful discussions. We acknowledge financial support from italian MIUR, through PRIN 2010-2011. We also acknowledge the CINECA award under the ISCRA initiative, for the availability of high performance computing resources and support. 
Part of the simulations were run on the LUCIA cluster at DiSAT, University of Insubria (Como).
\bibliographystyle{mn2e} 
\bibliography{MBHB1}

\begin{thebibliography}{37}
\expandafter\ifx\csname natexlab\endcsname\relax\def\natexlab#1{#1}\fi

\bibitem[{{Berger} \& {Colella}(1989)}]{berger89}
{Berger} M.~J., {Colella} P., 1989, Journal of Computational Physics, 82, 64

\bibitem[{{Berger} \& {Oliger}(1984)}]{berger84}
{Berger} M.~J., {Oliger} J., 1984, Journal of Computational Physics, 53, 484

\bibitem[{{Binney} \& {Tremaine}(2008)}]{binney08}
{Binney} J., {Tremaine} S., 2008, {Galactic Dynamics: Second Edition}.
  Princeton University Press

\bibitem[{{Blecha}, {Loeb} \& {Narayan}(2013){Blecha}, {Loeb}, \&
  {Narayan}}]{blecha13}
{Blecha} L., {Loeb} A., {Narayan} R., 2013, \mnras, 429, 2594

\bibitem[{{Bonoli}, {Mayer} \& {Callegari}(2014){Bonoli}, {Mayer}, \&
  {Callegari}}]{bonoli14}
{Bonoli} S., {Mayer} L., {Callegari} S., 2014, \mnras, 437, 1576

\bibitem[{{Callegari} {et~al}\mbox{.}(2011){Callegari}, {Kazantzidis}, {Mayer},
  {Colpi}, {Bellovary}, {Quinn}, \& {Wadsley}}]{callegari11}
{Callegari} S., {Kazantzidis} S., {Mayer} L., {Colpi} M., {Bellovary} J.~M.,
  {Quinn} T., {Wadsley} J., 2011, \apj, 729, 85

\bibitem[{{Callegari} {et~al}\mbox{.}(2009){Callegari}, {Mayer}, {Kazantzidis},
  {Colpi}, {Governato}, {Quinn}, \& {Wadsley}}]{callegari09}
{Callegari} S., {Mayer} L., {Kazantzidis} S., {Colpi} M., {Governato} F.,
  {Quinn} T., {Wadsley} J., 2009, \apjl, 696, L89

\bibitem[{{Chapon}, {Mayer} \& {Teyssier}(2013){Chapon}, {Mayer}, \&
  {Teyssier}}]{chapon13}
{Chapon} D., {Mayer} L., {Teyssier} R., 2013, \mnras, 429, 3114

\bibitem[{{Colpi} \& {Dotti}(2011)}]{colpi11}
{Colpi} M., {Dotti} M., 2011, Advanced Science Letters, 4, 181

\bibitem[{{Consortium} {et~al}\mbox{.}(2013){Consortium}, {:}, {Seoane},
  {Aoudia}, {Audley}, {Auger}, {Babak}, {Baker}, {Barausse}, {Barke}, {Bassan},
  {Beckmann}, {Benacquista}, {Bender}, {Berti}, {Bin{\'e}truy}, {Bogenstahl},
  {Bonvin}, {Bortoluzzi}, {Brause}, {Brossard}, {Buchman}, {Bykov}, {Camp},
  {Caprini}, {Cavalleri}, {Cerdonio}, {Ciani}, {Colpi}, {Congedo}, {Conklin},
  {Cornish}, {Danzmann}, {de Vine}, {DeBra}, {Dewi Freitag}, {Di Fiore}, {Diaz
  Aguilo}, {Diepholz}, {Dolesi}, {Dotti}, {Fern{\'a}ndez Barranco},
  {Ferraioli}, {Ferroni}, {Finetti}, {Fitzsimons}, {Gair}, {Galeazzi},
  {Garcia}, {Gerberding}, {Gesa}, {Giardini}, {Gibert}, {Grimani}, {Groot},
  {Guzman Cervantes}, {Haiman}, {Halloin}, {Heinzel}, {Hewitson}, {Hogan},
  {Holz}, {Hornstrup}, {Hoyland}, {Hoyle}, {Hueller}, {Hughes}, {Jetzer},
  {Kalogera}, {Karnesis}, {Kilic}, {Killow}, {Klipstein}, {Kochkina},
  {Korsakova}, {Krolak}, {Larson}, {Lieser}, {Littenberg}, {Livas}, {Lloro},
  {Mance}, {Madau}, {Maghami}, {Mahrdt}, {Marsh}, {Mateos}, {Mayer},
  {McClelland}, {McKenzie}, {McWilliams}, {Merkowitz}, {Miller}, {Mitryk},
  {Moerschell}, {Mohanty}, {Monsky}, {Mueller}, {M{\"u}ller}, {Nelemans},
  {Nicolodi}, {Nissanke}, {Nofrarias}, {Numata}, {Ohme}, {Otto},
  {Perreur-Lloyd}, {Petiteau}, {Phinney}, {Plagnol}, {Pollack}, {Porter},
  {Prat}, {Preston}, {Prince}, {Reiche}, {Richstone}, {Robertson}, {Rossi},
  {Rosswog}, {Rubbo}, {Ruiter}, {Sanjuan}, {Sathyaprakash}, {Schlamminger},
  {Schutz}, {Sch{\"u}tze}, {Sesana}, {Shaddock}, {Shah}, {Sheard}, {Sopuerta},
  {Spector}, {Spero}, {Stanga}, {Stebbins}, {Stede}, {Steier}, {Sumner}, {Sun},
  {Sutton}, {Tanaka}, {Tanner}, {Thorpe}, {Tr{\"o}bs}, {Tinto}, {Tu},
  {Vallisneri}, {Vetrugno}, {Vitale}, {Volonteri}, {Wand}, {Wang}, {Wanner},
  {Ward}, {Ware}, {Wass}, {Weber}, {Yu}, {Yunes}, \& {Zweifel}}]{lisa}
{Consortium} T.~e. {et~al.}, 2013, ArXiv e-prints

\bibitem[{{Daddi} {et~al}\mbox{.}(2010){Daddi}, {Elbaz}, {Walter}, {Bournaud},
  {Salmi}, {Carilli}, {Dannerbauer}, {Dickinson}, {Monaco}, \&
  {Riechers}}]{daddi10}
{Daddi} E. {et~al.}, 2010, \apjl, 714, L118

\bibitem[{{Davies}, {Miller} \& {Bellovary}(2011){Davies}, {Miller}, \&
  {Bellovary}}]{davies11}
{Davies} M.~B., {Miller} M.~C., {Bellovary} J.~M., 2011, \apjl, 740, L42

\bibitem[{{Di Matteo}, {Springel} \& {Hernquist}(2005){Di Matteo}, {Springel},
  \& {Hernquist}}]{dimatteo05}
{Di Matteo} T., {Springel} V., {Hernquist} L., 2005, \nat, 433, 604

\bibitem[{{Dotti} {et~al}\mbox{.}(2009){Dotti}, {Ruszkowski}, {Paredi},
  {Colpi}, {Volonteri}, \& {Haardt}}]{dotti09}
{Dotti} M., {Ruszkowski} M., {Paredi} L., {Colpi} M., {Volonteri} M., {Haardt}
  F., 2009, \mnras, 396, 1640

\bibitem[{{Dotti}, {Sesana} \& {Decarli}(2012){Dotti}, {Sesana}, \&
  {Decarli}}]{dotti12}
{Dotti} M., {Sesana} A., {Decarli} R., 2012, Advances in Astronomy, 2012

\bibitem[{{Dotti} {et~al}\mbox{.}(2010){Dotti}, {Volonteri}, {Perego}, {Colpi},
  {Ruszkowski}, \& {Haardt}}]{dotti10}
{Dotti} M., {Volonteri} M., {Perego} A., {Colpi} M., {Ruszkowski} M., {Haardt}
  F., 2010, \mnras, 402, 682

\bibitem[{{Dubois}, {Volonteri} \& {Silk}(2014){Dubois}, {Volonteri}, \&
  {Silk}}]{dubois14b}
{Dubois} Y., {Volonteri} M., {Silk} J., 2014, \mnras, 440, 1590

\bibitem[{{Dubois} {et~al}\mbox{.}(2014){Dubois}, {Volonteri}, {Silk},
  {Devriendt}, \& {Slyz}}]{dubois14a}
{Dubois} Y., {Volonteri} M., {Silk} J., {Devriendt} J., {Slyz} A., 2014,
  \mnras, 440, 2333

\bibitem[{{En{\ss}lin} {et~al}\mbox{.}(2007){En{\ss}lin}, {Pfrommer},
  {Springel}, \& {Jubelgas}}]{ensslin07}
{En{\ss}lin} T.~A., {Pfrommer} C., {Springel} V., {Jubelgas} M., 2007, \aap,
  473, 41

\bibitem[{{Escala} {et~al}\mbox{.}(2005){Escala}, {Larson}, {Coppi}, \&
  {Mardones}}]{escala05}
{Escala} A., {Larson} R.~B., {Coppi} P.~S., {Mardones} D., 2005, \apj, 630, 152

\bibitem[{{Ferrara}, {Haardt} \& {Salvaterra}(2013){Ferrara}, {Haardt}, \&
  {Salvaterra}}]{ferrara13}
{Ferrara} A., {Haardt} F., {Salvaterra} R., 2013, \mnras, 434, 2600

\bibitem[{{Fiacconi} {et~al}\mbox{.}(2013){Fiacconi}, {Mayer}, {Ro{\v s}kar},
  \& {Colpi}}]{fiacconi13}
{Fiacconi} D., {Mayer} L., {Ro{\v s}kar} R., {Colpi} M., 2013, \apjl, 777, L14

\bibitem[{{Gabor} \& {Bournaud}(2013)}]{gabor13}
{Gabor} J.~M., {Bournaud} F., 2013, ArXiv e-prints

\bibitem[{{Genzel} {et~al}\mbox{.}(2010){Genzel}, {Tacconi}, {Gracia-Carpio},
  {Sternberg}, {Cooper}, {Shapiro}, {Bolatto}, {Bouch{\'e}}, {Bournaud},
  {Burkert}, {Combes}, {Comerford}, {Cox}, {Davis}, {Schreiber},
  {Garcia-Burillo}, {Lutz}, {Naab}, {Neri}, {Omont}, {Shapley}, \&
  {Weiner}}]{genzel10}
{Genzel} R. {et~al.}, 2010, \mnras, 407, 2091

\bibitem[{{Hernquist}(1990)}]{hernquist90}
{Hernquist} L., 1990, \apj, 356, 359

\bibitem[{{Hobbs} {et~al}\mbox{.}(2010){Hobbs}, {Archibald}, {Arzoumanian},
  {Backer}, {Bailes}, {Bhat}, {Burgay}, {Burke-Spolaor}, {Champion}, {Cognard},
  {Coles}, {Cordes}, {Demorest}, {Desvignes}, {Ferdman}, {Finn}, {Freire},
  {Gonzalez}, {Hessels}, {Hotan}, {Janssen}, {Jenet}, {Jessner}, {Jordan},
  {Kaspi}, {Kramer}, {Kondratiev}, {Lazio}, {Lazaridis}, {Lee}, {Levin},
  {Lommen}, {Lorimer}, {Lynch}, {Lyne}, {Manchester}, {McLaughlin}, {Nice},
  {Oslowski}, {Pilia}, {Possenti}, {Purver}, {Ransom}, {Reynolds}, {Sanidas},
  {Sarkissian}, {Sesana}, {Shannon}, {Siemens}, {Stairs}, {Stappers},
  {Stinebring}, {Theureau}, {van Haasteren}, {van Straten}, {Verbiest},
  {Yardley}, \& {You}}]{PTA}
{Hobbs} G. {et~al.}, 2010, Classical and Quantum Gravity, 27, 084013

\bibitem[{{Hopkins} {et~al}\mbox{.}(2012){Hopkins}, {Hernquist}, {Hayward}, \&
  {Narayanan}}]{hopkins12}
{Hopkins} P.~F., {Hernquist} L., {Hayward} C.~C., {Narayanan} D., 2012, \mnras,
  425, 1121

\bibitem[{{Kormendy}(2013)}]{kormendy13}
{Kormendy} J., 2013, {Secular Evolution in Disk Galaxies}, {Falc{\'o}n-Barroso}
  J., {Knapen} J.~H., eds., p.~1

\bibitem[{Lupi {et~al}\mbox{.}(2014)Lupi, Colpi, Devecchi, Galanti, \&
  Volonteri}]{lupi14}
Lupi A., Colpi M., Devecchi B., Galanti G., Volonteri M., 2014, Monthly Notices
  of the Royal Astronomical Society, 442, 3616

\bibitem[{{Maio} {et~al}\mbox{.}(2013){Maio}, {Dotti}, {Petkova}, {Perego}, \&
  {Volonteri}}]{maio13}
{Maio} U., {Dotti} M., {Petkova} M., {Perego} A., {Volonteri} M., 2013, \apj,
  767, 37

\bibitem[{{Mayer} {et~al}\mbox{.}(2010){Mayer}, {Kazantzidis}, {Escala}, \&
  {Callegari}}]{mayer10}
{Mayer} L., {Kazantzidis} S., {Escala} A., {Callegari} S., 2010, \nat, 466,
  1082

\bibitem[{{Rasera} \& {Teyssier}(2006)}]{rasera06}
{Rasera} Y., {Teyssier} R., 2006, \aap, 445, 1

\bibitem[{{Sesana} {et~al}\mbox{.}(2014){Sesana}, {Barausse}, {Dotti}, \&
  {Rossi}}]{Sesana14}
{Sesana} A., {Barausse} E., {Dotti} M., {Rossi} E.~M., 2014, ArXiv e-prints

\bibitem[{{Springel}(2005)}]{Springel05}
{Springel} V., 2005, \mnras, 364, 1105

\bibitem[{{Teyssier}(2002)}]{teyssier02}
{Teyssier} R., 2002, \aap, 385, 337

\bibitem[{{Teyssier} {et~al}\mbox{.}(2013){Teyssier}, {Pontzen}, {Dubois}, \&
  {Read}}]{teyssier13}
{Teyssier} R., {Pontzen} A., {Dubois} Y., {Read} J.~I., 2013, \mnras, 429, 3068

\bibitem[{{Van Wassenhove} {et~al}\mbox{.}(2012){Van Wassenhove}, {Volonteri},
  {Mayer}, {Dotti}, {Bellovary}, \& {Callegari}}]{vanwassenhove12}
{Van Wassenhove} S., {Volonteri} M., {Mayer} L., {Dotti} M., {Bellovary} J.,
  {Callegari} S., 2012, \apjl, 748, L7

\end{thebibliography}
\label{lastpage}
\end{document}